\documentclass[journal, 10pt]{IEEEtran}

\usepackage[noadjust]{cite}

\ifCLASSINFOpdf
   \usepackage[pdftex]{graphicx}
\else
\fi
\usepackage[cmex10]{amsmath}
\usepackage{array}

\usepackage{mdwmath}
\usepackage{mdwtab}

\usepackage{eqparbox}

\usepackage[hang,raggedright,normalsize,tight]{subfigure}

\usepackage{caption}
\usepackage{url}

\usepackage{amsfonts}
\usepackage{tabularx}
\usepackage{tikz, pgfplots}
\usetikzlibrary{arrows}
\usetikzlibrary{patterns}
\usetikzlibrary{plotmarks}
\usetikzlibrary{arrows,positioning} 
\usetikzlibrary{decorations.markings}
\tikzset{
    >=stealth',
    punkt/.style={
           rectangle,
           rounded corners,
           draw=black, very thick,
           text width=6.5em,
           minimum height=2em,
           text centered},
    pil/.style={
           ->,
           thick,
           shorten <=2pt,
           shorten >=2pt,}
    pir/.style={
           <-,
           thick,
           shorten <=2pt,
           shorten >=2pt,}
}
\definecolor{KeynoteRed}{rgb}{.678,.051, .051}
\definecolor{KeynoteBlue}{rgb}{0.008, 0.443, 0.60}
\definecolor{DarkBlue}{rgb}{0, 0, 0.6}
\definecolor{KeynoteLightblue}{rgb}{.635, .914, .973}
\definecolor{KeynoteYellow}{rgb}{0.859, 0.584, 0.212}
\definecolor{KeynoteSlate}{rgb}{0.239, 0.271, 0.322}
\definecolor{KeynoteGray}{rgb}{0.498, 0.529, 0.529}
\definecolor{KeynoteTextGray}{rgb}{0.325, 0.325, 0.325}
\definecolor{KeynoteLightGray}{rgb}{0.706, 0.706, 0.706}
\definecolor{KeynoteBlueGray}{rgb}{0.471, 0.533, 0.620}
\definecolor{KeynoteGreen}{rgb}{0.18, 0.50, 0.09}
\definecolor{KeynotePurple}{rgb}{0.18, 0.50, 0.09}
\definecolor{ECEpurple}{rgb}{.169, .18, .455}
\definecolor{ECEcyan}{rgb}{.41, .62, .72}
\definecolor{ECEgray}{rgb}{.788, .827, .859}
\definecolor{ECEblueGray}{rgb}{61.2, 70.6, 70.6}
\definecolor{ECEblueGray}{rgb}{61.2, 70.6, 70.6}
\definecolor{RiceBlue}{rgb}{0, .14, .41}

\definecolor{D}{rgb}{0.859, 0.584, 0.212}
\definecolor{DA}{rgb}{0.008, 0.443, 0.60} 
\definecolor{DC}{rgb}{.678,.051, .051} 
\definecolor{DCA}{rgb}{.10, .10, .3} 

\newcommand{\note}[1]{{\color{red} {\scriptsize \emph{#1}}}}

\newcommand{\revision}[1]{{\color{black} {#1}}}
\newcommand{\revOption}[1]{{\color{black}{{\footnotesize #1}}}}

\newcommand{\delaySpread}{\sigma_{\rm \tau}}
\newcommand{\DRR}{{\sf DRR}}

\newcommand{\SSINRFD}{{\sf SSINR}^{\rm FD}_{\rm Uplink}}

\newcommand{\SNRUP}{{\sf SNR}^{\rm HD}_{\rm Uplink}}

\newcommand{\Ex}{\mathbb{E}}

\newcommand{\RFDUP}{R_{\rm Uplink}^{\rm FD}}

\newcommand{\RHDUP}{R_{\rm Uplink}^{\rm HD}}

\newcommand{\recip}{two-node}
\newcommand{\nonrecip}{three-node}
\newcommand{\Recip}{Two-node}
\newcommand{\NonRecip}{Three-node}
\newcommand{\degree}{^\circ}
\newcommand{\dirBigA}{I}
\newcommand{\dirBigB}{II}
\newcommand{\dirSmall}{III}
\newcommand{\omniBig}{IV}
\newcommand{\omniSmall}{V}
\newcommand{\avgAlpha}{\bar \alpha^{(\mathsf{mech})}_{\rm P}}
\newtheorem{theorem}{Theorem}
\newtheorem{result}{Result}

\newif\ifIncludeMechanismDiscussion
\newif\ifIncludeNoAnalogResult
\newif\ifShowCBWTables
\newif\ifGenerateFigures
\newif\ifLongAnalyticalDiscussion
\newif\ifSimpleSuppression
\newif\ifLongDesignImplication
\newif\ifLongFreSelectIllustration

\begin{document}

\title{Passive Self-Interference Suppression for Full-Duplex Infrastructure Nodes}

\author{Evan~Everett,
		Achaleshwar~Sahai, 
        and~Ashutosh~Sabharwal
\thanks{The authors would like to thank the Antenna and Wireless Systems Branch of NASA's Johnson Space Center for both technical advice and the use of the Antenna Test Facility. In particular, the authors would like to recognize Michael Khayat for his guidance in experiment design, and Wayne Cope for his EM measurement expertise. The authors also thank Jingwen Bai and Guarav Patel for their crucial assistance.}

\thanks{This work was partially supported by NSF Grants CNS 0923479, CNS 1012921 and CNS 1161596 and NSF Graduate Research Fellowship 0940902.}
\thanks{E. Everett, A. Sahai and A. Sabharwal are with the Department of Electrical and Computer Engineering, Rice University, Houston, TX, 77005 USA. e-mail: \{evan.everett, as27, ashu\}@rice.edu.}
}

\maketitle

\begin{abstract}
\label{sec:abstract}
Recent research results have demonstrated the feasibility of full-duplex wireless communication for short-range links. Although the focus of the previous works has been \emph{active} cancellation of the self-interference signal, a majority of the overall self-interference suppression is often due to \emph{passive} suppression, i.e., isolation of the transmit and receive antennas. We present a measurement-based study of the capabilities and limitations of three key mechanisms for passive self-interference suppression: directional isolation, absorptive shielding, and cross-polarization. The study demonstrates that more than 70~dB of passive suppression can be achieved in certain environments, but also establishes two results on the limitations of passive suppression: (1) environmental reflections limit the amount of passive suppression that can be achieved, and (2) passive suppression, in general, increases the frequency selectivity of the residual self-interference signal. These results suggest two design implications: (1) deployments of full-duplex infrastructure nodes should minimize near-antenna reflectors, and (2) active cancellation in concatenation with passive suppression should employ higher-order filters or per-subcarrier cancellation.  
\end{abstract}

\IEEEpeerreviewmaketitle

\IncludeMechanismDiscussionfalse

\IncludeNoAnalogResultfalse

\LongAnalyticalDiscussionfalse

\SimpleSuppressiontrue

\GenerateFigurestrue

\LongDesignImplicationfalse

\LongFreSelectIllustrationfalse

\section{Introduction}
\label{sec:intro}



Conventional wireless devices operate in half-duplex mode to avoid the high-powered self-interference that is generated when transmission and reception coexist in time and frequency. 
Recent results \cite{Bliss07SimultTX_RX, Khandani2010FDPatent, Radunovic:2009aa, Duarte10FullDuplex, Choi10FullDuplex, Jain2011RealTimeFD, Duarte11FullDuplex, Sahai11FullDuplex,  Khojastepour11AntennaCancellation, Aryafar12MIDU, Duarte2012FullDuplexWiFi, Duarte12Thesis}, however, have demonstrated the feasibility of short-range full-duplex wireless communication by mitigating self-interference through a combination of passive suppression and active cancellation  of the self-interference. Passive suppression is any technique to electromagnetically isolate the transmit and receive antennas, while active suppression is any technique to exploit a node's knowledge of its own transmit signal to cancel the self-interference.
Active cancellation, which has been studied extensively in the literature, functions by injecting a cancellation waveform into the receive signal path to null the self-interference. The cancellation waveform can be injected over a transmission line as in \cite{Duarte10FullDuplex, Jain2011RealTimeFD, Sahai11FullDuplex, Duarte2012FullDuplexWiFi, Duarte12Thesis} or over the air using extra antennas as in \cite{ Bliss07SimultTX_RX, Choi10FullDuplex, Khojastepour11AntennaCancellation, Aryafar12MIDU}. Passive suppression, however, has been studied only sparsely. With the aim to develop a better understanding of the capabilities and limitations of passive self-interference suppression, this paper presents a measurement-based characterization of passive suppression mechanisms for full-duplex infrastructure nodes.


Our motivation for studying passive suppression stems from the observation that passive suppression accounts for a large portion of the total self-interference suppression in existing full-duplex designs. For example, of the 85~dB of self-interference suppression achieved in the full-duplex design in \cite{Duarte2012FullDuplexWiFi}, 65~dB is attributed to passive suppression and only 20~dB to active cancellation. 
Moreover, recent characterizations of active cancellation~\cite{SahaiPhaseNoiseDraft, Day12FDRelay, Day12FDMIMO} have shown that the amount of self-interference that can be actively cancelled is limited by fundamental radio impairments such as phase noise~\cite{SahaiPhaseNoiseDraft} and limited dynamic range~\cite{Day12FDRelay, Day12FDMIMO}, but it is not known whether passive suppression will also encounter fundamental bottlenecks and what those bottlenecks may be. 
Thus, an opportunity exists to improve full-duplex performance through a richer understanding of capabilities and limitations of passive suppression mechanisms. 

\subsection{Contribution}
Our first (and primary) contribution is an experimental characterization of passive self-interference suppression methods for full-duplex infrastructure nodes. The characterization focuses on three key passive suppression mechanisms: (1) directional isolation -- the use of directional antennas such that the gain of the transmit antenna is low in the direction of the receive and visa versa, (2) absorptive shielding --  the use of lossy materials to attenuate the self-interference, and (3) cross-polarization -- the use of transmit and receive antennas in orthogonal polarization states. 
The characterization has two main goals, (a) to quantify the performance and limitations of the three mechanisms and combinations among the three mechanisms, and (b) to study the effect of passive suppression on the characteristics of the self-interference channel.
To accomplish these goals, the three mechanisms were evaluated both in a best-case environment of an anechoic chamber, and a worst-case environment of a highly reflective room with metal walls. In each environment the self-interference channel was measured for several different antenna configurations both with and without the three passive suppression mechanisms in place. 
The characterization establishes \ifIncludeNoAnalogResult three \else two \fi main results which are summarized below.

\ifIncludeNoAnalogResult
\emph{Result~\ref{result:noAnalog}:} In a low-reflection environment, more that $70$~dB of passive self-interference suppression can be reliably achieved by employing the three mechanisms in tandem. At typical WiFi ranges, $70$~dB of passive suppression is sufficient to force the self-interference to a power level commensurate to the received signal from another node, thus preventing the self-interference from dominating the ADC dynamic range and enabling the active cancelation of residual self-interference to be implemented purely digitally. 
\fi

\emph{Result~\ref{result:reflections}:}
Environmental reflections present a fundamental bottleneck in the amount of passive suppression that can be achieved.  Comparing the measurements in the anechoic chamber to those in the reflective room, it is observed that although the passive mechanisms are quite effective in suppressing the direct path between antennas, none can predictably attenuate reflected self-interference paths. 
The design implication is that, in order to achieve high-levels of passive suppression, deployers of full-duplex infrastructure should seek to minimize nearby reflectors. 

\emph{Result~\ref{result:coherence}:}
Passive self-interference suppression, in general, causes a decrease in the coherence bandwidth of the residual self-interference channel. Since the passive suppression operates on the direct path between transmit and receive antennas, but not necessarily reflected paths, passive suppression can transition the self-interference channel from a predominantly line-of-sight channel to a multi-path dominated channel, where frequency selectivity occurs due to constructive and destructive combinations of reflected paths. The design implication is that when active cancellation is employed in concatenation with passive suppression, the active canceler should be designed to handle a frequency-selective self-interference signal.   

As a second contribution, we present a prototype full-duplex infrastructure node, implemented in the WARPLab \cite{WARPLab} framework, that combines the passive suppression mechanisms studied in the above characterization with the existing active cancellation mechanisms studied in~\cite{Duarte11FullDuplex,Duarte2012FullDuplexWiFi}.
We evaluate the performance of this design in an outdoor full-duplex uplink, and see that total self-interference suppression exceeding 90~dB can be achieved, enabling the prototype to outperform a comparable half-duplex link even at ranges exceeding 100~m.

The above two results, when combined with previous works, paint a more complete picture of self-interference mitigation. Passive suppression is the first line-of-defense against self-interference, but is effective only against the direct path. 
The residual multi-path self-interference signal can be further suppressed by a per-subcarrier active cancellation strategy \cite{Duarte11FullDuplex, Duarte2012FullDuplexWiFi} that is robust to frequency selectivity. Unfortunately, active cancelation also encounters bottlenecks due to RF impairments such as phase noise \cite{SahaiPhaseNoiseDraft} and limited dynamic range \cite{Day12FDMIMO,Day12FDRelay}. Because both passive and active suppression encounter limitations, it is important that full-duplex designs reach their full potential in both the active \emph{and} passive suppression regimes. 

 
\subsection{Prior Art}
Many of the full-duplex designs in the literature have employed one or more passive suppression method. 
In the  full-duplex designs of \cite{Bliss07SimultTX_RX, Duarte10FullDuplex, Choi10FullDuplex, Jain2011RealTimeFD, Duarte11FullDuplex} the only passive suppression mechanism utilized is physical separation of transmit and receive antennas, i.e., free-space path loss. 
In \cite{Sahai11FullDuplex, Duarte2012FullDuplexWiFi} heuristic strategies for laptop antenna placement are studied including ``device-induced path loss,'' in which transmit and receive antennas are shielded by laptop hardware such as the screen. Such techniques were used in \cite{Duarte2012FullDuplexWiFi} to achieve as much as $65$~dB of passive suppression. 
Passive suppression for full-duplex relay systems in which the transmit and receive antennas are not collocated, and large amounts of path loss can be leveraged, are studied in \cite{Slingsby95AntennaIsolation, Anderson04AntennaIsoloation, Haneda10RelaySuppression}.
In~\cite{Slingsby95AntennaIsolation, Anderson04AntennaIsoloation} directional isolation and 4-5~m of antenna separation are leveraged to achieve 75-90~dB of passive suppression. 
In \cite{Haneda10RelaySuppression} an outdoor-to-indoor relay systems is studied, in which directional isolation and cross-polarization are leveraged. In a deployment where the antennas are on opposite sides of a wall, 48~dB of passive suppression is observed, and when the indoor antenna is moved into the interior of the building, benefitting from large path loss, 60-80~dB of passive suppression is observed. 
In contrast to \cite{Slingsby95AntennaIsolation, Anderson04AntennaIsoloation, Haneda10RelaySuppression}, our study focuses on more compact full-duplex deployments in which the antenna separation does not exceed 50~cm. 
Moreover, our study, which includes directional isolation, cross-polarization and absorptive shielding, is to our knowledge the first comprehensive study of passive self-interference suppression for wireless full-duplex systems. In particular, none of the previous designs have quantified the impact of environmental reflections on the performance of passive suppression. 

\subsection{Organization of Paper}
The organization of the rest of the paper is as follows. 
\ifIncludeMechanismDiscussion Section~\ref{sec:mechanisms} we briefly describe the three passive suppression mechanisms characterized in our study.\fi In Section~\ref{sec:setup} we describe  the experiment design in detail. In Section~\ref{sec:characterization} we present the outcomes of the experiment, with special focus on the two main results listed above.
\revision{
In Section~\ref{sec:analysis} we analyze the impact of passive suppression on capacity.
}
In Section~\ref{sec:outdoor} we present a prototype and evaluate the performance of combined passive and active suppression in a wideband OFDM full-duplex physical layer.
\revision{
In Section~\ref{sec:users} we have some system-level discussion on implementing passive suppression at infrastructure nodes.
}
We conclude in Section~\ref{sec:discussion}.

\ifIncludeMechanismDiscussion
	\section{Mechanisms for Passive Self-Interference Suppression}
\label{sec:mechanisms}
Below is a brief description of three mechanisms characterized in our study: directional isolation, absorptive shielding, and cross-polarization. For a more detailed background on these mechanisms and how they can be leveraged in practical designs, please see~\cite{Everett12Thesis}. 

\subsection{Directional Isolation}
Full-duplex operation requires extra antenna and RF resources that space-constrained mobile devices may not accommodate in the near future.
Therefore the first applications of full-duplex may be scenarios where an infrastructure node can leverage full-duplex to enhance spectral efficiency when communicating with half-duplex mobile clients.
Many such scenarios possess \emph{directional diversity},  meaning that the direction of the infrastructure nodes's transmission is different from the direction of reception. One example of such a scenario is a full-duplex relay \cite{Haneda10RelaySuppression,riihonen11}, and another is a sectorized full-duplex access~point in which the access~point transmits a downlink signal to a client in one sector while simultaneously receiving an uplink signal from another client in another sector in the same band \cite{Everett12Thesis} (\note{Also cite Jingwen if her paper is ready}) as shown in figure~\ref{fig:mechanisms}.
In such \emph{directional diversity} scenarios, directional antennas can be employed such that the transmit energy can be directed \emph{away} from the receiving antenna as shown in Figure~\ref{fig:mechanisms}.  Similarly the receive pattern is pointed \emph{away} from the transmitting antenna. Thus the coupling between transmitting and receiving antennas is greatly reduced via directional isolation. 

\begin{figure}
		\centering
	\ifCLASSOPTIONonecolumn
		\includegraphics[width=0.5\textwidth]{./DocGraphics/mechanisms.pdf}
	\else
		\includegraphics[width=0.4\textwidth]{./DocGraphics/mechanisms.pdf}
	\fi
	\caption{Shielding via RF absorver and cross-polarization via dual-polarized antennas}
	\label{fig:mechanisms}
\end{figure}

\subsection{Absorptive Shielding}
In infrastructure nodes, sufficient real-estate is likely available to strucurally shield the receiving antenna(s) from the transmitting antenna(s). 
Metal conductors are excellent RF shields, but a conductor in the near-field of an antenna will couple with the antenna, detuning it (i.e. making it a less efficient radiator) and unpredictably distorting the coverage pattern. Moreover, even if the conductor is in the far-field of the antenna, it will reflect the energy back into the desired coverage area producing an interference pattern. 
Materials are readily available, however, that \emph{absorb} incident RF energy rather than reflecting it, thus suppressing the self-interference without distorting coverage.  
RF absorber material is most-commonly implemented as a carbon-loaded foam that coverts incident RF energy to heat via ohmic losses \cite{Tong}. See \cite{Everett12Thesis} and references therein for a detailed discussion of appropriate choice of RF absorber for passive self-interference suppression. 

\subsection{Cross-polarization}
The \emph{polarization} of an electromagnetic wave is the dimension in which the electric field vector is oscillating \cite{Stutzman}. Two antennas are said to be ``co-polarized'' if they transmit and receive waves with parallel polarization and ``cross-polarized'' if they transmit/receive waves with orthogonal polarization.
In addition to time, frequency, and space, polarization is one of the fundamental dimensions of an electromagnetic signal \cite{TsePoon11DOF_Polarization}, but it is the dimension that is often least exploited in today's wireless network since polarization is dependent on antenna orientation, which can be hard to predict with mobile devices. 
The self-interference channel of a full-duplex node, however, is between antennas located on the same device, and thus with fixed relative orientation. It therefore makes sense to exploit cross-polarization for suppressing self-interference as shown in Figure~\ref{fig:mechanisms}. 
\fi

\section{Experiment Design}
\label{sec:setup}

The goal of the experiments presented in this paper is to characterize the performance of three key mechanisms for passive self-interference suppression in full-duplex infrastructure nodes, and to study how the three mechanisms affect the self-interference channel. To accomplish this goal, 
a network analyzer was used to characterize the self-interference channel between transmitting and receiving antennas as the three passive suppression mechanisms were applied to several different antenna configurations both in a reflective environment and in a nonreflective environment.  The following is a detailed description of the experiment. 


\subsection{Experiment Setup}
We begin by describing the directional antenna configurations studied, and then describe how absorptive shielding and cross-polarization are leveraged within in each configuration.

\begin{figure*}[tbhp]
\centering
\subfiguretopcapfalse
\renewcommand{\subcapsize}{\footnotesize}
\subfigure[Configuration\hspace{.5em}\dirBigA: $90\degree$~beamwidth antennas, $90\degree$~beam separation, 50~cm~antenna separation.]{
	\centering
	\ifCLASSOPTIONonecolumn
		\includegraphics[natwidth=399,natheight=260,width=0.28\textwidth, trim=0 10pt 0 0]{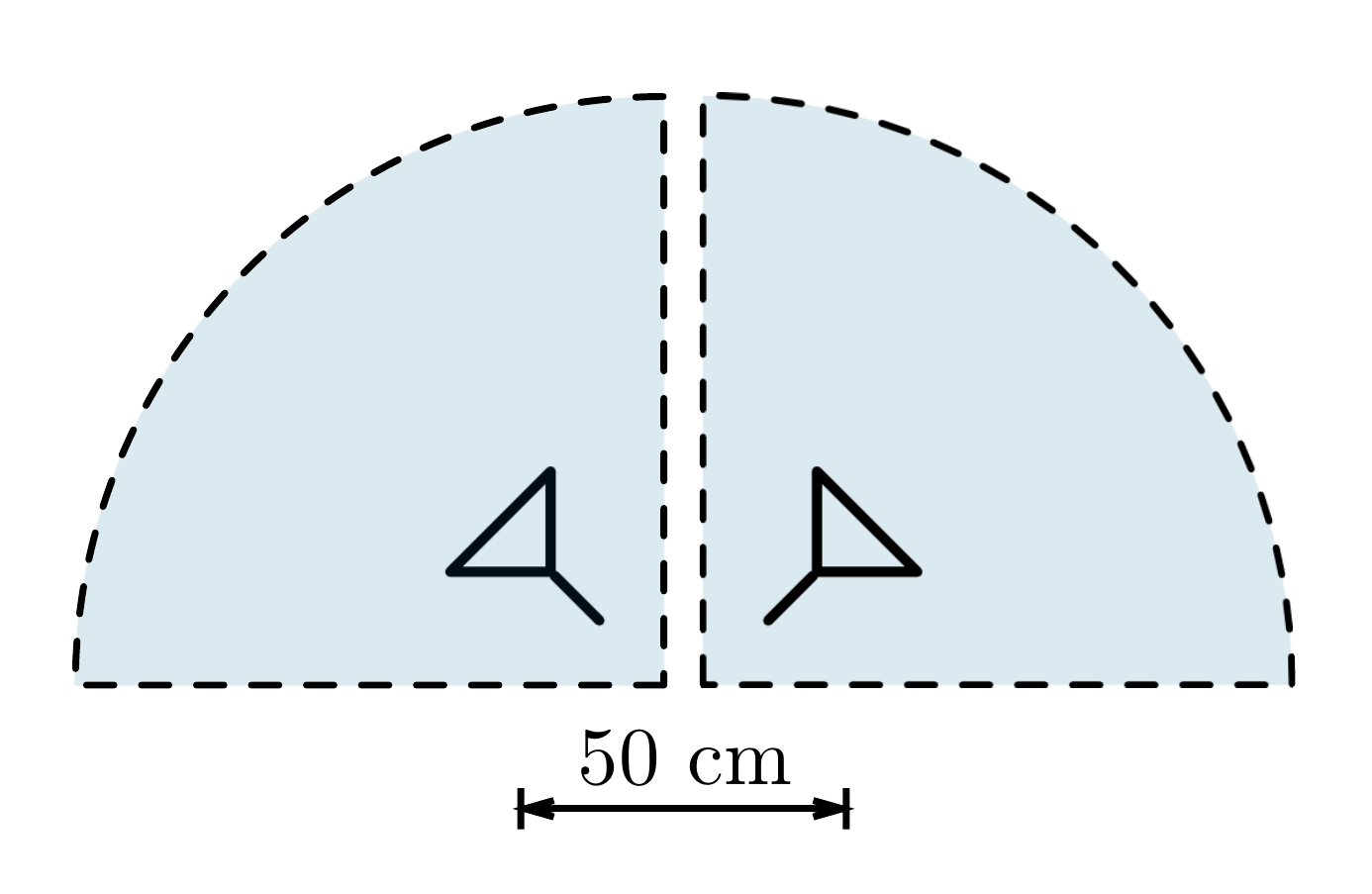}
		\hspace{10pt}
	\else
		\includegraphics[natwidth=399,natheight=260,width=0.28\textwidth, trim=0 10pt 0 0]{./DocGraphics/Config_90_50-eps-converted-to.pdf}
	\fi
	\label{fig:config_90_50}
}
\subfigure[Configuration\hspace{.5em}\dirBigB: $90\degree$~beamwidth antennas, $60\degree$~beam separation, 50~cm~antenna separation.]{
	\centering
	\ifCLASSOPTIONonecolumn
		\includegraphics[natwidth=398,natheight=251,width=0.28\textwidth, trim=0 10pt 0 0]{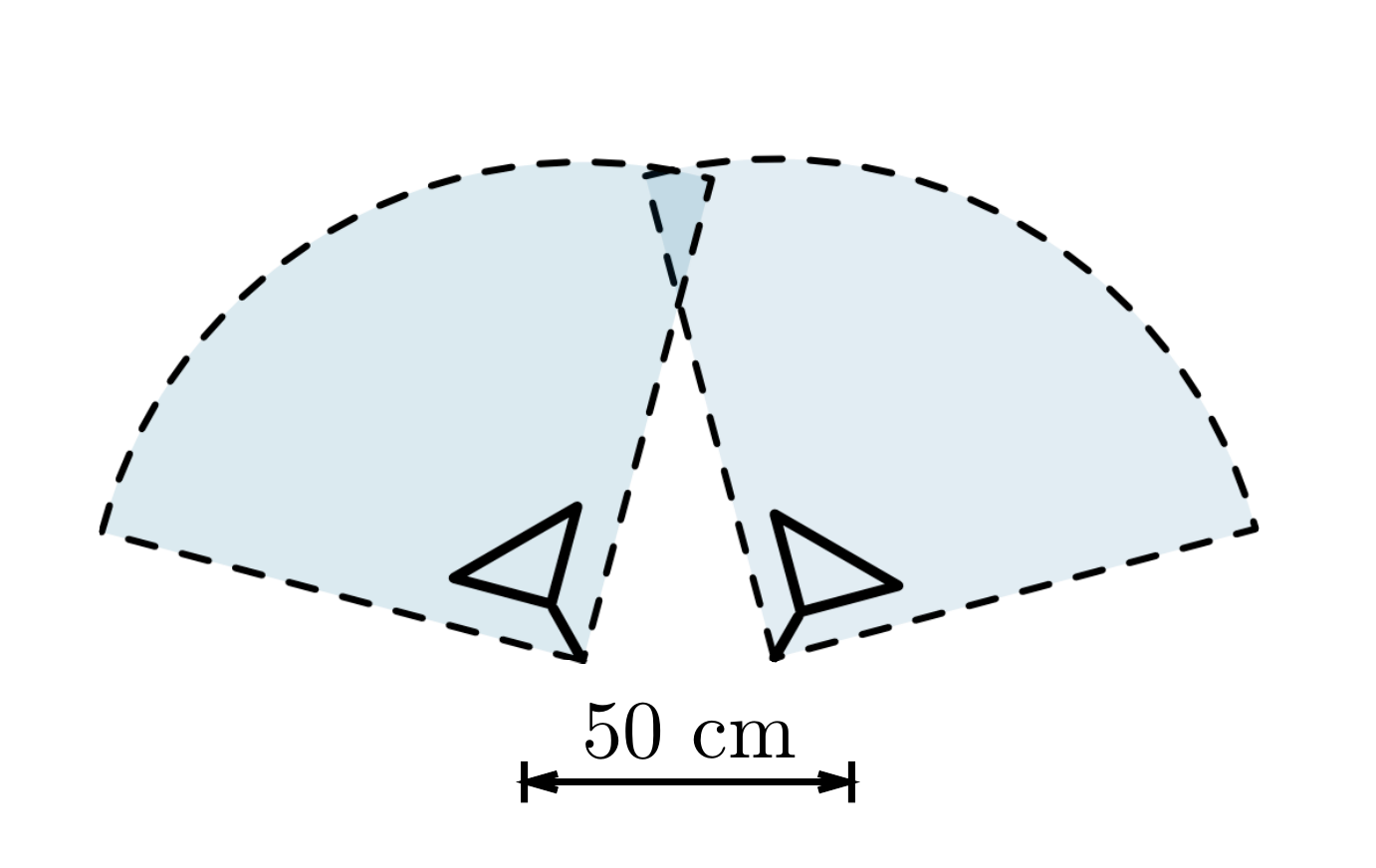}
	\else
		\includegraphics[natwidth=398,natheight=251, width=0.28\textwidth, trim=0 10pt 0 0]{./DocGraphics/Config_60_50-eps-converted-to.pdf}
	\fi
	\label{fig:config_60_50}
}
\subfigure[Configuration\hspace{.5em}\dirSmall: $90\degree$~beamwidth antennas, $60\degree$~beam separation, 35~cm~antenna separation.]{	\label{fig:config_60_35}
	\centering
	\ifCLASSOPTIONonecolumn
		\includegraphics[natwidth=398,natheight=251, width=0.28\textwidth, trim=0 10pt 0 0]{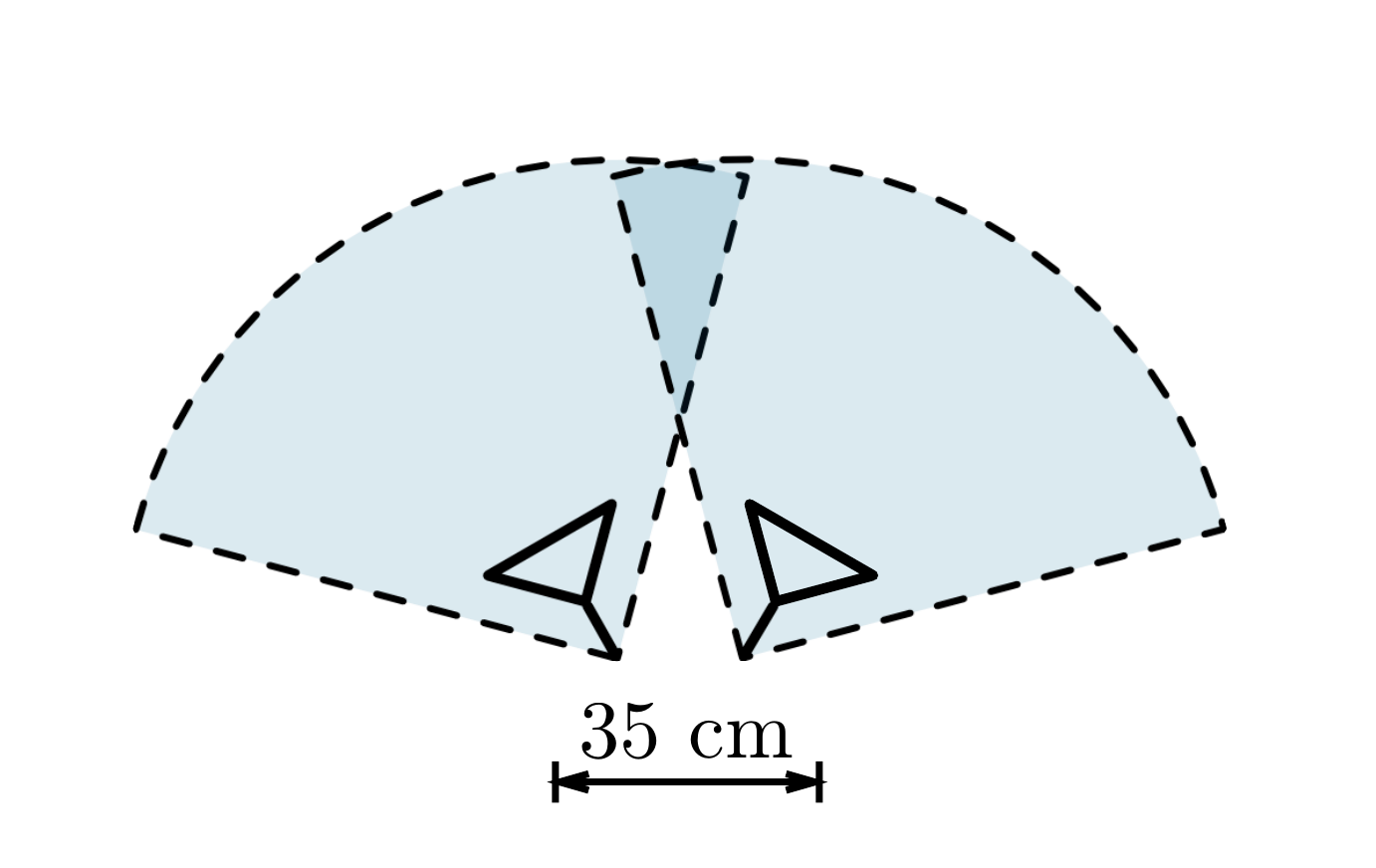}
	\else
		\includegraphics[width=0.28\textwidth, trim=0 10pt 0 0]{./DocGraphics/Config_60_35-eps-converted-to.pdf}
	\fi
}
\\
\subfigure[Configuration~\omniBig: Omnidirectional~antennas, 50~cm~antenna separation.]{\label{fig:config_omni_50}
	\centering
	\ifCLASSOPTIONonecolumn
		\hspace{15pt}
		\includegraphics[natwidth=399,natheight=251,width=0.28\textwidth, trim=0 30pt 0 10pt]{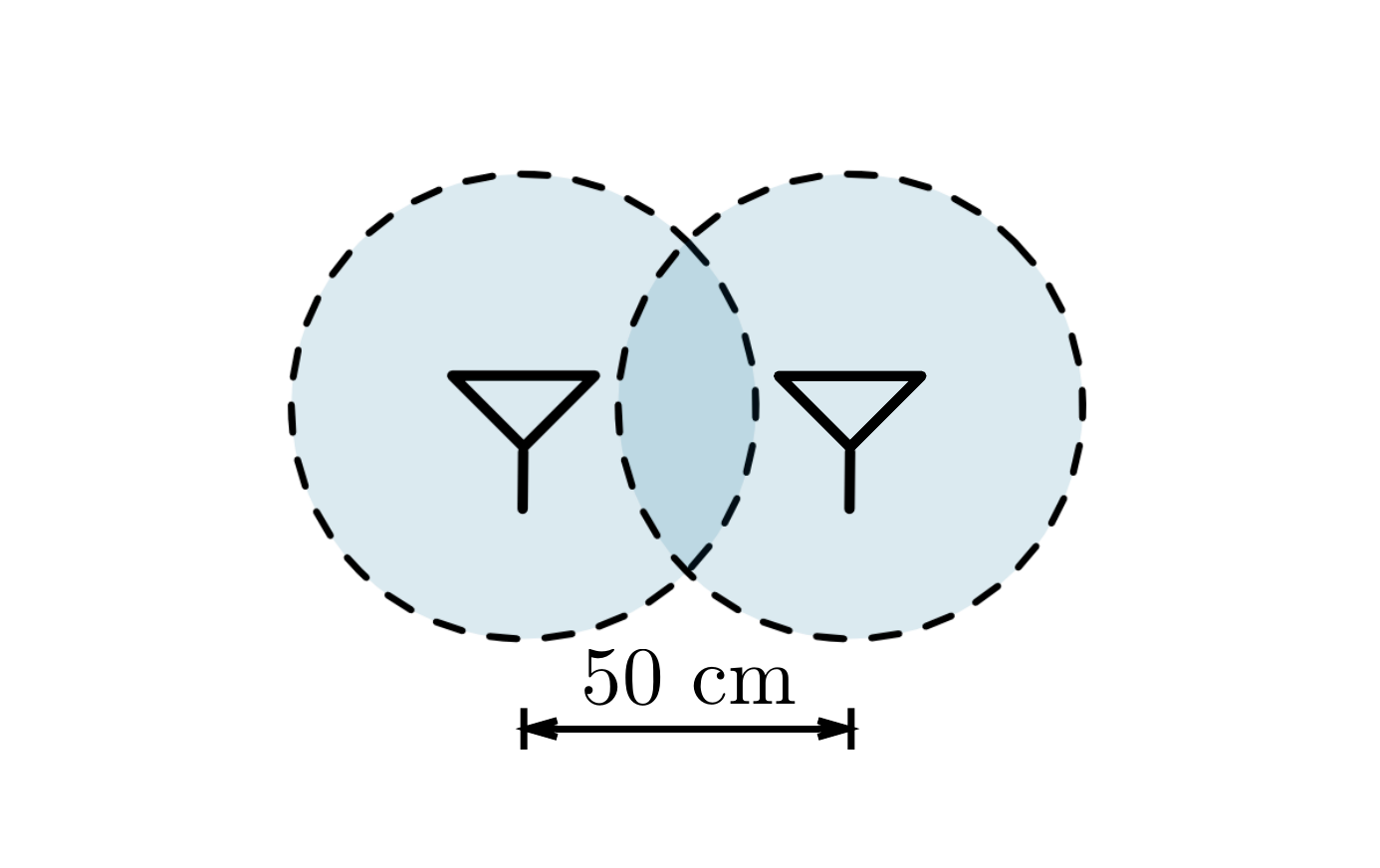}
		\hspace{15pt}
	\else
		\includegraphics[natwidth=399,natheight=251,width=0.28\textwidth, trim=0 35pt 0 20pt]{./DocGraphics/Config_Omni_50-eps-converted-to.pdf}
	\fi
	}
\hspace{20pt}
\subfigure[Configuration~\omniSmall: Omnidirectional~antennas, 35~cm~antenna separation.]{\label{fig:config_omni_35}
	\centering
	\ifCLASSOPTIONonecolumn
		\hspace{15pt}
		\includegraphics[natwidth=398,natheight=251,width=0.28\textwidth, trim=0 30pt 0 10pt]{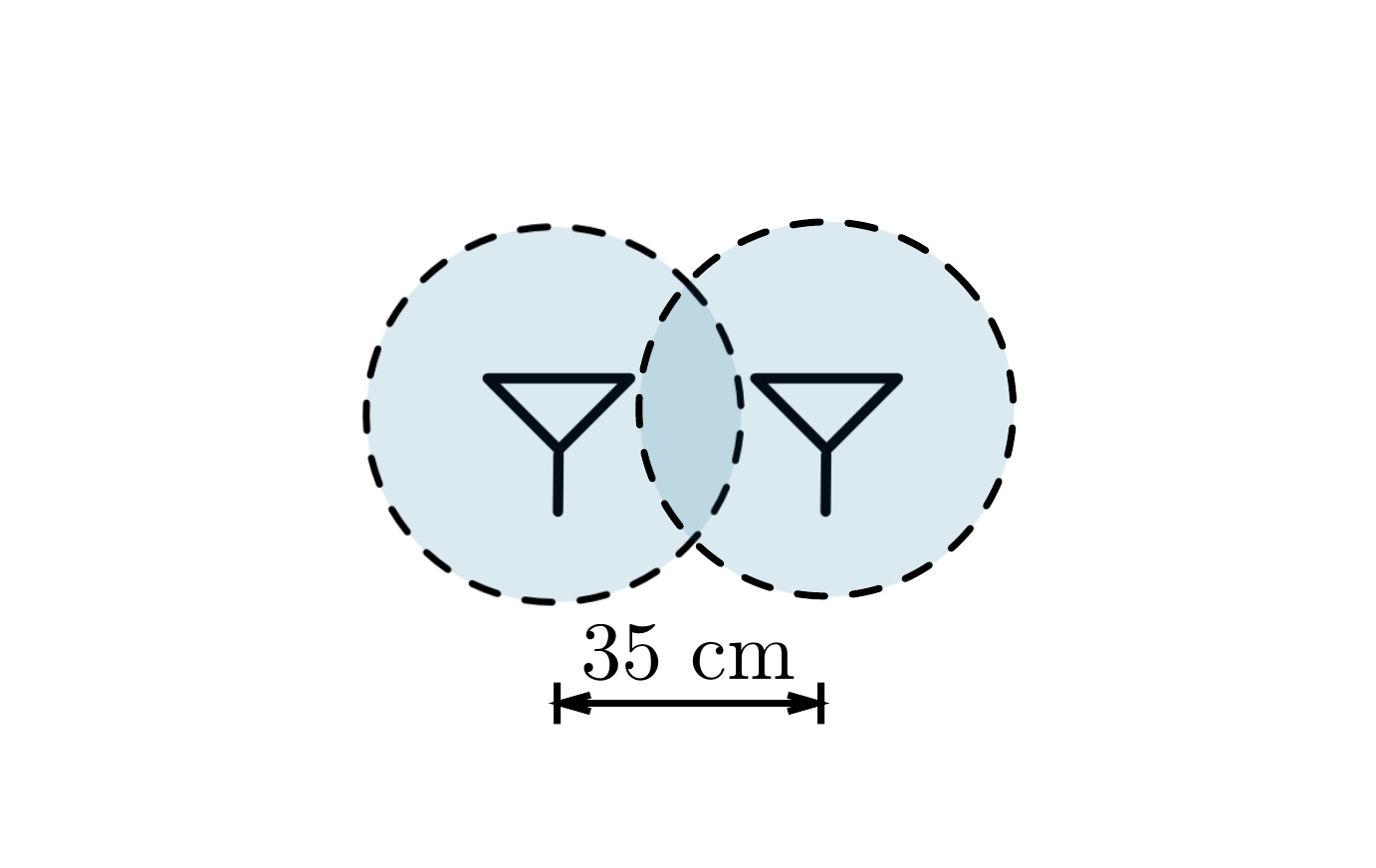}
		\hspace{15pt}
	\else
		\includegraphics[natwidth=398,natheight=251,width=0.28\textwidth, trim=0 35pt 0 20pt]{./DocGraphics/Config_Omni_35-eps-converted-to.pdf}
	\fi
}
\caption{Five architectures evaluated in the passive suppression characterization. We do not designate one antennas as the transmitter and one as receiver as these roles exchange.
}
\label{fig:configurations}
\end{figure*}

\subsubsection{Implementation of Directional Isolation}
The five infrastructure antenna configurations in Figure~\ref{fig:configurations} were used to study passive self-interference suppression via directional isolation. The configurations differ in the directionality of the antennas and in their relative orientation.
Configuration~\dirBigA, illustrated in Figure~\ref{fig:config_90_50}, uses $90\degree$ beamwidth\footnote{Throughout the paper when we use the term ``beamwidth'' we mean the 3 dB beamwidth -- the span of angles for which the gain of the antenna is within 3 dB of its maximum gain. The conical antennas patterns in Figure~\ref{fig:configurations} are only figurative: actual patterns have a gradual rolloff in gain and also have side lobes.} antennas which are pointed with $90\degree$ beam separation, so that there is little overlap in the coverage patterns.  Configuration~\dirBigB, shown in Fig~\ref{fig:config_60_50}, also  uses $90\degree$ beamwidth antennas, but with a beam separation of $60\degree$, so that there is $30\degree$ overlap between the coverage zones. In both Configurations~\dirBigA~and~\dirBigB\ the separation between antennas is 50~cm. Configuration~\dirSmall, shown in Figure~\ref{fig:config_60_35}, is identical to Configuration~\dirBigB, except that the separation between antennas is scaled down to 35~cm. The $90\degree$ beamwidth antennas used in Configurations~\dirBigA,~\dirBigB,~and~\dirSmall\ are HG2414DP-090 panel antennas from L-com \cite{Lcom90} designed for outdoor sectorized WiFi access~points. 
For comparing to cases without directional isolation, Configurations~\omniBig~and~\omniSmall, shown in Figures~\ref{fig:config_omni_50}~and~\ref{fig:config_omni_35}, were also studied. The 50~cm separation between  antennas in Configuration~\omniBig~is the same as in Configurations~\dirBigA~and~\dirBigB, the 35cm separation in Configuration~\omniSmall~is the same as in Configuration~\dirSmall. 
The omnidirectional antennas used in Configurations~\omniBig~and~\omniSmall\ are HGV-2406 outdoor WiFi antennas from L-com \cite{LcomOmni}.

\subsubsection{Implementation of Absorptive Shielding}
Absorptive shielding is realized by placing a slab of RF absorber material between the two antennas as illustrated in Figure~\ref{fig:mechanisms}. The material used in our experiments is Eccosorb AN-79 \cite{Absorber} free-space RF absorber. AN-79 is a broadband, tapered loading absorber made from polyurethane foam impregnated with a carbon gradient. It is a 4.25 inch slab that can be cut to fit the application. The manufacturer's data sheet indicates that AN-79 can provide up to 25 dB of absorption.

\ifIncludeMechanismDiscussion
\else
\begin{figure}
		\centering
	\ifCLASSOPTIONonecolumn
		\includegraphics[natwidth=398,natheight=215,width=0.4\textwidth]{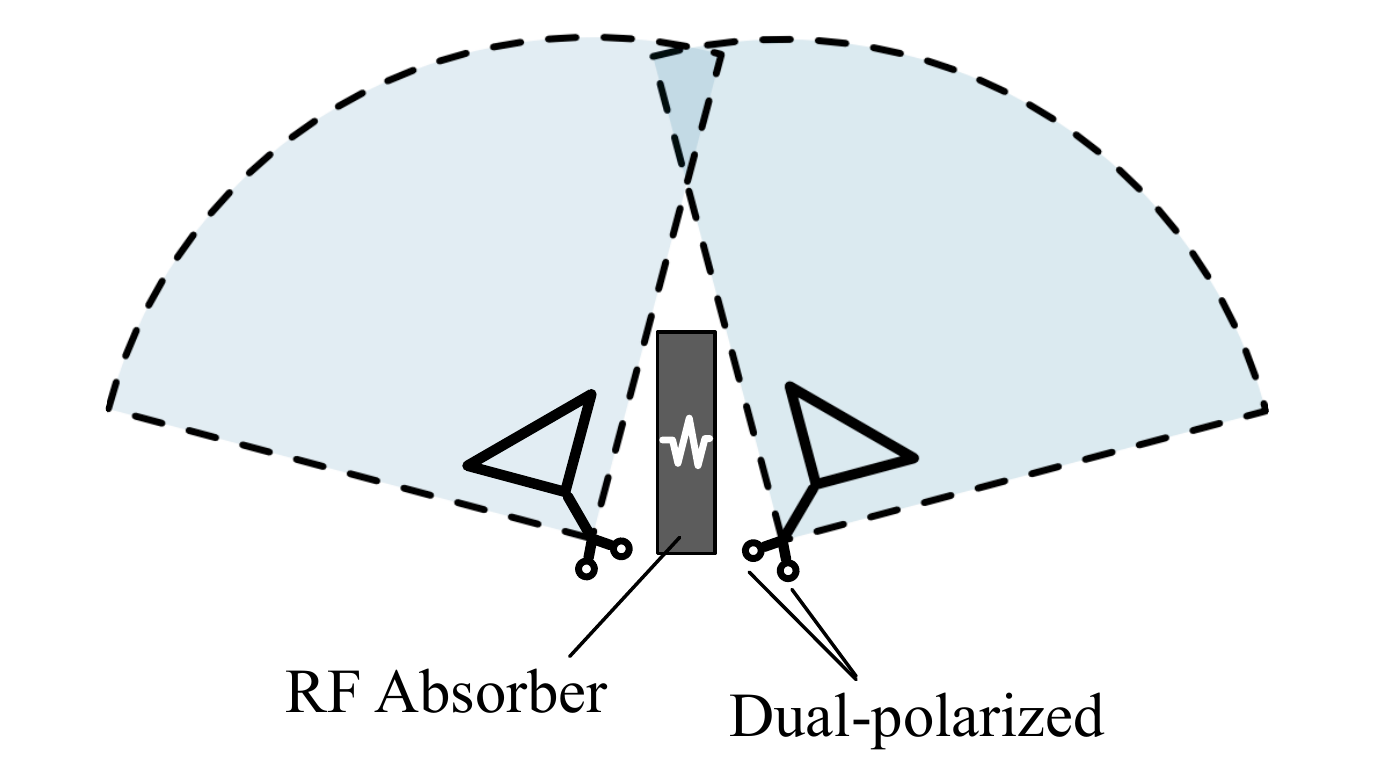}
	\else
		\includegraphics[natwidth=398,natheight=215,width=0.3\textwidth]{./DocGraphics/PolAbs-eps-converted-to.pdf}
	\fi
	\caption{Shielding via RF absorver and cross-polarization via dual-polarized antennas.
	\label{fig:mechanisms}}
\end{figure}
\fi

\subsubsection{Implementation of Cross-polarization}
The L-Com HGV-2406 antennas used in the directional-antenna configurations are dual-polarized as indicated in Figure~\ref{fig:mechanisms}, which means they have two ports: one excites a horizontally polarized mode, and the other a vertically polarized mode. 
By measuring the coupling between both co-polarized and cross-polarized ports, we can study the impact of cross-polarization on the self-interference channel.
%
Most commercial omnidirectional antennas support only a single polarization mode,\footnote{Most commercial omnidirectional antennas are implemented as an electric dipole or an electric monopole, which only support a vertically polarized mode. Directional antennas, however, are commonly implemented as circular or rectangular patches, which can support two modes: one with vertical polarization and the other with horizontal polarization. 
}
and such is the case for the L-Com HGV-2406 antennas used in the omnidirectional configurations of Figure~\ref{fig:configurations}, which only support vertical polarization. Thus passive suppression via cross-polarization was not studied in the omnidirectional configurations.

\subsection{Environments}
In order for the measurements to be as repeatable as possible, the first stage of the passive suppression characterization was performed in a shielded anechoic chamber to minimize reflections and interference from other sources. 
After the measurements were performed in the anechoic chamber, they were repeated in a highly reflective room to observe the effect of environmental reflections. The room used was roughly 30 ft. $\times$ 60 ft. with metal walls and a low metal ceiling, intended to represent a worst~case reflection environment for practical deployments. 

\subsection{Measurements and Metrics}

\subsubsection{Measurement Procedure}
%

For notational convenience let the variable,
\begin{equation}
	\mathsf{mech}\ \in\ \{{\sf \emptyset,  D,A,C,DA,DC,AC,DCA}\},
\end{equation}
denote a combination of passive suppression mechanisms, where $\emptyset$ denotes no passive suppression, $\mathsf{D}$ denotes directional isolation, $\mathsf{A}$ denotes absorptive shielding, and $\mathsf{C}$ denotes cross-polarization. Note that the antenna configuration constrains which combinations are possible. 
Since directional antennas are used in Configurations~\dirBigA,~\dirBigB,~and~\dirSmall, the possible values of $\mathsf{mech}$ are $\{\mathsf{D, DA, DC, DCA}\}$ for these configurations. Similarly the possible values for $\mathsf{mech}$ in Configurations~\omniBig~and~\omniSmall\ are $\{\mathsf{\emptyset, A}\}$, since the antennas are omnidirectional and not dual-polarized. 
Therefore, for each of the directional-antenna configurations shown in Figure~\ref{fig:configurations}, the self-interference channel between antennas was measured for the following combinations of passive self-interference suppression mechanisms: directional isolation only ($\mathsf{mech = D}$), directional isolation plus absorptive shielding ($\mathsf{mech = DA}$), directional isolation plus cross-polarization ($\mathsf{mech = DC}$), and directional isolation plus absorptive shielding plus cross-polarization ($\mathsf{mech = DCA}$). For the omnidirectional configurations,  the combination of mechanisms measured were no suppression ($\mathsf{mech = \emptyset}$) and absorptive shielding ($\mathsf{mech = A}$).
%
For each combination of applied suppression mechanisms, $\mathsf{mech}$, we directly measured the frequency response $H^{(\mathsf{mech})}(f)$ of the self-interference channel between the two antennas using an Agilent E8364B \cite{PNA_E8364B} general-purpose network analyzer (PNA). The channel response, $H^{(\mathsf{mech})}(f)$, was obtained by connecting the transmit and receive antenna ports to the network analyzer, configured to record the S-parameters of a two-port network at 20,000 uniformly spaced frequency components in the WiFi ISM band, $f\in [2.40, 2.48]$~GHz. Each frequency sample was averaged over $20$ measurements to reduce noise. 

\subsubsection{Metrics}
\ifSimpleSuppression
Let $\bar \alpha^{(\mathsf{mech})}_{\rm P}$ denote the amount of passive suppression achieved by the combination of mechanisms $\mathsf{mech}$. We define $\bar \alpha^{(\mathsf{mech})}_{\rm P}$ to be ratio of the transmit power to the power of the self-interference incident on the receiver, which is the inverse of the power in the frequency response, $H^{(\mathsf{mech})}(f)$. Thus we compute  $\bar \alpha^{(\mathsf{mech})}_{\rm P}$ from the measured frequency response according to
\begin{equation}
\bar \alpha^{(\mathsf{mech})}_{\rm P} = \left( \frac{1}{N} \sum_{f=f_{min}}^{f_{max}} \left|H^{(\mathsf{mech})}(f)\right|^2 \right)^{-1}. %
\end{equation} 
\else
We denote the amount of passive self-interference suppression that a combination of mechanisms, $\mathsf{mech}$, achieves at frequency, $f$, as
\begin{equation}
\alpha^{(\mathsf{mech})}_{\rm P}(f) \equiv \frac{1}{\left|H^{(\mathsf{mech})}(f)\right|^2}.
\end{equation}
We will often refer to $\alpha^{(\mathsf{mech})}_{\rm P}$ as the ``passive suppression strength,'' the ``passive suppression'', or just the ``suppression''.  
A figure of merit that will be used to compare the performance of different suppression mechanisms is the average passive suppression over the frequency response, 
\begin{equation}
\bar \alpha^{(\mathsf{mech})}_{\rm P} = \left( \frac{1}{N} \sum_{f=f_{min}}^{f_{max}} \left|H^{(\mathsf{mech})}(f)\right|^2 \right)^{-1}. 
\end{equation}
\fi

In addition to studying the amount of suppression that different mechanisms attain, we also want to study how passive suppression affects the frequency-selectivity of the self-interference channel.  
The metric most often used to quantify the frequency-selectivity of a wireless channel is the coherence bandwidth, $B_C$. 
The definition of coherence bandwidth we adopt is the range of frequencies over which the channel response is at least $90\%$ correlated \cite{GoldsmithWirelessComm}. 
The coherence bandwidth is difficult to measure directly, but can computed from the power delay profile of the channel, $P(\tau)$, which can be easily measured. In our study, the power-delay profile is obtained by taking the inverse discrete time Fourrier transform (DTFT) of the measured frequency response. The RMS delay spread of the channel, $\delaySpread$, can be computed from the power delay profile, which is related to the $90\%$ coherence bandwidth via the approximation \cite{LeeWirelessCommBook} 
\begin{equation}
\label{eq:CBW}
B_C \approx \frac{0.02}{\delaySpread}.
\end{equation}
Let $P_0(\tau)$ denote the normalized power delay profile, that is
$
	P_0(\tau) = \frac{P(\tau)}{\int_0^{\infty}P(\tau) d\tau}.
$
The mean delay, $\mu_{\tau}$ is given by
	$\mu_{\tau} = \int_0^{\infty} \tau P_0(\tau) d\tau,$
and the RMS delay spread, $\delaySpread$, can then be computed as \cite{GoldsmithWirelessComm}.
\begin{equation}
	\sigma_{\tau} = \sqrt{\int_{0}^\infty (\tau - \mu_{\tau})^2 P_0(\tau) d\tau}. 
\end{equation}
Taking $\delaySpread$, computed as shown above from the measured power delay profile, and plugging into Equation~(\ref{eq:CBW}) allows approximation of the coherence bandwidth from the empirical channel response. 
If we think of the normalized power delay, $P_0(\tau)$, as the ``distribution'' on the delay $\tau$ then the above equations are easy to interpret.

\section{Experiment Results}
\label{sec:characterization}

\renewcommand{\arraystretch}{1.5}
\renewcommand{\subcapsize}{\normalsize}
\subfiguretopcaptrue
\begin{table*}[htdp]
\caption{Passive Suppression Measurements \label{tab:Suppression}}
\begin{center}
%
\subfigure[\normalsize{\underline{Anechoic Chamber}} \label{tab:DirChamber}]{
	\begin{tabular}{|l||>{\centering}p{1.3cm}|>{\centering}p{1.3cm}|>{\centering}p{1.3cm}|}
\hline
	Applied Mechanism ($\mathsf{mech}$) & $\avgAlpha$ Config~\dirBigA & $\avgAlpha$ Config~\dirBigB & $\avgAlpha$ Config~\dirSmall \tabularnewline \hline \hline
	Directional $(\sf D)$ &45.3~dB \revOption{(0.12~dB)}&  39.7~dB \revOption{(0.10~dB)}& 45.1~dB \revOption{(0.12~dB)}\tabularnewline \hline
	${\sf D}$ + Absorber $(\sf DA)$ & 55.4~dB \revOption{(0.22~dB)}&  50.9~dB \revOption{(0.16~dB)}& 55.9~dB \revOption{(0.24~dB)}\tabularnewline \hline
	${\sf D}$ + Cross-polarization $(\sf DC)$ & 63.9~dB  \revOption{(0.51~dB)}&  58.3~dB \revOption{(0.28~dB)}& 56.9~dB \revOption{(0.26~dB)}\tabularnewline \hline
	${\sf D}$ + Cross-pol + Absorber $(\sf DCA)$ & 73.8~dB \revOption{(1.5~dB)}& 72.5~dB \revOption{(1.2~dB)} & 72.4~dB \revOption{(1.2~dB)}\tabularnewline \hline
	\end{tabular}
	}
\subfigure[\normalsize{\underline{Reflective Room}} \label{tab:DirRoom}]{
	\begin{tabular}{|>{\centering}p{1.3cm}|>{\centering}p{1.3cm}|>{\centering}p{1.3cm}|}
\hline
	 $\avgAlpha$ Config~\dirBigA & $\avgAlpha$ Config~\dirBigB & $\avgAlpha$ Config~\dirSmall \tabularnewline \hline \hline
	 37.7~dB \revOption{(0.10~dB)}& 36.0~dB \revOption{(0.09~dB)}& 37.1~dB \revOption{(0.09~dB)}\tabularnewline \hline
	 39.4~dB \revOption{(0.10~dB)}& 38.1~dB \revOption{(0.10~dB)}& 37.4~dB \revOption{(0.10~dB)}\tabularnewline \hline
	 45.1~dB \revOption{(0.12~dB)}& 44.9~dB \revOption{(~0.12dB)}& 44.4~dB \revOption{(0.11~dB)}\tabularnewline \hline
	 45.9~dB \revOption{(0.12~dB)}& 44.9~dB \revOption{(0.12~dB)}& 45.7~dB \revOption{(0.12~dB)}\tabularnewline \hline
	\end{tabular}
	}
\\ 
\subfigure[\normalsize{\underline{Anechoic Chamber}} \label{tab:OmniChamber}]{
	\begin{tabular}{|l||>{\centering}p{1.3cm}|>{\centering}p{1.3cm}|}
	\hline
	Applied Mechanism ($\mathsf{mech}$)& $\avgAlpha$ Config~\omniBig & $\avgAlpha$ Config~\omniSmall \tabularnewline \hline\hline
	None $(\sf \emptyset)$ & 27.9~dB \revOption{(0.08~dB)}& 24.5~dB \revOption{(0.08~dB)}\tabularnewline \hline
	Absorber $(\sf A)$ & 48.0~dB \revOption{(0.14~dB)}& 48.2~dB \revOption{(0.14~dB)}\tabularnewline \hline
	\end{tabular}
	}
\subfigure[\normalsize{\underline{Reflective Room}} \label{tab:OmniRoom}]{
	\begin{tabular}{|>{\centering}p{1.3cm}|>{\centering}p{1.3cm}|}
	\hline
	$\avgAlpha$ Config~\omniBig & $\avgAlpha$ Config~\omniSmall \tabularnewline \hline \hline
 	27.6~dB \revOption{(0.08~dB)}& 25.1~dB \revOption{(0.08~dB)}\tabularnewline \hline
	40.1~dB \revOption{(0.10~dB)}& 41.1~dB \revOption{(0.10~dB)}\tabularnewline \hline
	\end{tabular}
	}
\hspace{20pt}
\subfigure[\normalsize{\underline{Legend}}]{
	\begin{tabular}{|>{\centering}p{1.8cm}|}
	\hline
	$\avgAlpha$ Configuration \tabularnewline \hline
	Value \newline \revOption{(Uncertainty)}\tabularnewline \hline 
	\end{tabular}
	}
\end{center}
\end{table*} 

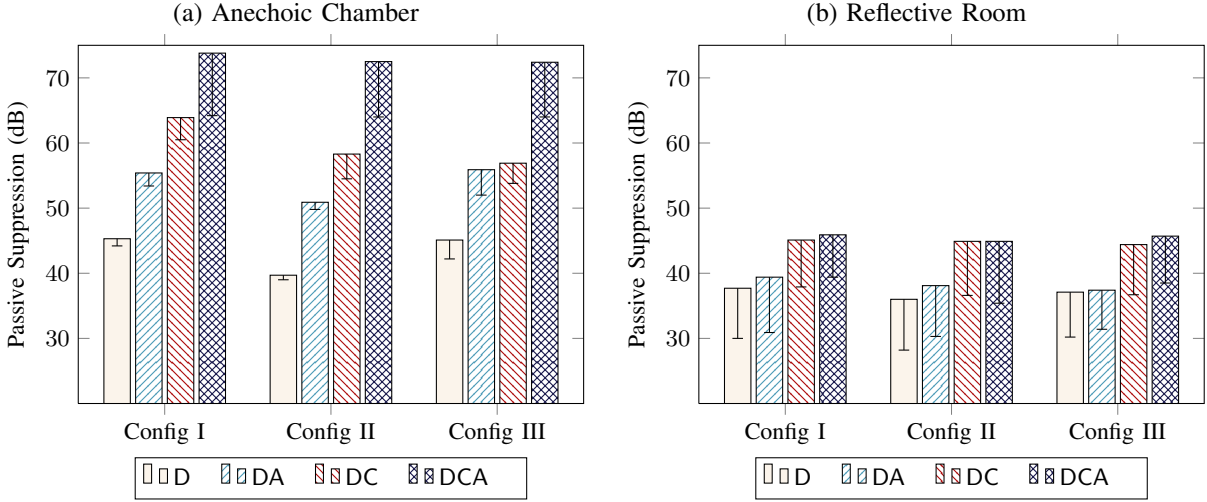
\begin{figure*}[htbp]
\begin{center}
\subfigure[Anechoic Chamber \label{fig:suppressionChamber}]{
	{\small
\begin{tikzpicture} 
\begin{axis}[
width=8.3cm,
ylabel=Passive Suppression (dB), 
ylabel near ticks, 
enlarge x limits= .26,
height=0.35\textwidth,
legend style={at={(0.5,-0.15)},area legend,
anchor=north,legend columns=-1}, 
ybar = 2pt,
bar width=10pt, 
xlabel near ticks,
ylabel near ticks,
ymin=20, ymax=75,
ytick = {30,40,...,70},
symbolic x coords={Config I,Config II,Config III},
xtick = data,
]
\addplot[fill=D!10] [error bars/.cd, y dir=minus,y explicit]
	 coordinates {
	 	(Config I, 45.3) +- (1.2,1.1)
		(Config II, 39.7) +- (0.9,0.7)
		(Config III, 45.1) +- (3.2,2.9) 
};
\addplot[pattern=north east lines, pattern color = DA!70] [error bars/.cd, y dir=minus,y explicit] 
	coordinates {
		(Config I, 55.4) +- (3.3,2.0) 
		(Config II, 50.9) +- (1.4,1.1)
		(Config III, 55.9) +- (6.6,3.9)
};
\addplot[pattern=north west lines, pattern color = KeynoteRed] [error bars/.cd, y dir=minus,y explicit] 
	coordinates {
		(Config I, 63.9) +- (14.1,3.4)
		(Config II, 58.3) +- (9.1,3.8)
		(Config III, 56.9) +- (4.8,3.1)
};		  
\addplot[pattern=crosshatch, pattern color = DCA] [error bars/.cd, y dir=minus,y explicit] 
	coordinates {
		(Config I, 73.8) +- (39.4,9.6)
		(Config II, 72.5) +- (31.0,8.5)
		(Config III, 72.4) +- (35.5,8.4)
};
\legend{${\sf D\ \ \ }$, ${\sf DA\ \ \ }$, ${\sf DC\ \ \ }$, ${\sf DCA\ \ \ }$}
\end{axis} 
\end{tikzpicture}
}
}\subfigure[Reflective Room \label{fig:suppressionRoom}]{
	{\small
\begin{tikzpicture} 
\begin{axis}[
width=8.3cm,
height=0.35\textwidth,
ylabel=Passive Suppression (dB), 
ylabel near ticks, 
enlarge x limits= .26,
legend style={at={(0.5,-0.15)},
anchor=north,legend columns=-1}, 
ybar = 2pt,
ylabel near ticks,
xlabel near ticks,
bar width=10pt, 
ymin=20, ymax=75,
ytick = {30,40,...,70},
symbolic x coords={Config I,Config II,Config III},
xtick = data,
]
\addplot[fill=D!10][error bars/.cd, y dir=minus,y explicit]
	 coordinates {
	 	(Config I, 37.7) +- (27.8,7.7) 
		(Config II, 36.0) +- (36.7,7.8) 
		(Config III, 37.1) +- (42.9,6.9)  
};
\addplot[pattern=north east lines, pattern color = DA!70] [error bars/.cd, y dir=minus,y explicit] 
	coordinates {
		(Config I, 39.4) +- (50.6,8.5)
		(Config II, 38.1) +- (35.0,7.8) 
		(Config III, 37.4) +- (36.8,6.0) 
};
\addplot[pattern=north west lines, pattern color = KeynoteRed] [error bars/.cd, y dir=minus,y explicit] 
	coordinates {
		(Config I, 45.1) +- (40.4,7.2)
		(Config II, 44.9) +- (38.6,8.3)
		(Config III, 44.4) +- (34.9,7.7) 
};		  
\addplot[pattern=crosshatch, pattern color = DCA] [error bars/.cd, y dir=minus,y explicit] 
	coordinates {
		(Config I, 45.9) +- (37.3,6.5)
		(Config II, 44.9) +- (32.1,9.5) 
		(Config III, 45.7) +- (35.4,7.2)
};
\legend{${\sf D\ \ \ }$, ${\sf DA\ \ \ }$, ${\sf DC\ \ \ }$, ${\sf DCA\ \ \ }$}
\end{axis} 
\end{tikzpicture}
}
}
\end{center}
\caption{Passive suppression measurements with directional isolation ($\mathsf{D}$), directional isolation + absorptive shielding ($\mathsf{DA}$), directional isolation + cross-polarization ($\mathsf{DC}$), and directional isolation + cross-polarization shielding + cross-polarization ($\mathsf{DCA}$). The worst-case (minimum) suppression over the frequency band is indicated by the error bar. 
\label{fig:Suppression}}
\end{figure*}

Table~\ref{tab:Suppression} shows the passive suppression, $\avgAlpha$, measured for all of the configurations, both in the anechoic chamber and in the reflective room. Table~\ref{tab:DirChamber} contains the suppression measurements for the directional-antenna configurations in the anechoic chamber, and Table~\ref{tab:DirRoom} compares these measurements to those made in the reflective room. Similarly, Table~\ref{tab:OmniChamber} shows the passive suppression measurements for the omnidirectional configurations in the anechoic chamber, and Table~\ref{tab:OmniRoom} shows the measurements for the omnidirectional configurations is the reflective room. 
\revision{In parenthesis under each entry is the uncertainty in each measurement due to network analyzer noise and calibration error\footnote{\revision{The uncertainty values in Table~\ref{tab:Suppression} account for noise and calibration bias of the measurement equipment. The extent to which the fabrication of the antenna configuration is repeatable will also affect the uncertainty, but is very difficult to quantify. In the case of the anechoic chamber measurements, the reported uncertainties do not account for possible imperfections in the anechoic chamber, which is also hard to empirically quantify.}}.
See Appendix~\ref{sec:uncertainty} for an explanation of how these uncertainty values are obtained.} 
For easier visualization, we also plot the directional-configuration data shown in Tables~\ref{tab:DirChamber}~and~\ref{tab:DirRoom} in Figure~\ref{fig:Suppression}. \revision{
To indicate how far the worst-case passive suppression at any particular frequency can differ from average passive suppression over the frequency band, the minimum suppression over the frequency band is indicated by a one-sided ``error bar'' (and does \emph{not} correspond to the measurement uncertainty given in Table~\ref{tab:Suppression})}.

First consider the passive suppression achieved for Configuration~\dirBigA\ in the low-reflection environment of the anechoic chamber, shown in the first cluster of Figure~\ref{fig:suppressionChamber}.
Looking at the first bar in Figure~\ref{fig:suppressionChamber} we see that when Configuration~\dirBigA\ is used without any absorptive shielding or cross-polarization, the passive suppression is 45~dB. 
Comparing to the omnidirectional Configuration~\omniBig\ in the first column of Table~\ref{tab:OmniChamber}, we see that the directional isolation ($\mathsf{D}$) that comes from using the $90\degree$ beamwidth antennas with $90\degree$ beam separation provides around 17~dB of passive suppression in addition to the 28~dB of path loss between antennas. 
Adding absorptive shielding ($\mathsf{DA}$) gives an extra 10~dB of suppression for 55~dB total suppression. Similarly, adding cross-polarization ($\mathsf{DC}$) gives an extra 18~dB of suppression for a total suppression of 66~dB. When all the mechanisms are applied in tandem, directional isolation absorptive shielding, and cross-polarization together ($\mathsf{DCA}$) yield 74~dB total suppression. A similar trend is followed for the other directional-antenna configurations shown in Figure~\ref{fig:suppressionChamber}.\revision{\footnote{\revision{We see in the first row of Tables~\ref{tab:DirChamber}~and~Tables~\ref{tab:DirRoom} that the suppression with directional isolation ($\mathsf{D}$) is greater for Configuration~\dirBigB\ than Configuration~\dirSmall, both in the anechoic chamber and in the reflective room. This is surprising because the antennas are closer in Configuration~\dirSmall\ than Configuration~\dirBigB. We believe this is an artifact of the near-field coupling between the specific antennas used, not evidence of a general trend.}}}
\ifIncludeMechanismDiscussion
\else
We see in Figure~\ref{fig:suppressionChamber} that, when in the low-reflection environment of the anechoic chamber, leveraging directional isolation, cross-polarization, absorptive shielding, in tandem ($\mathsf{DCA}$) provides more than 70~dB of passive self-interference suppression in all three directional-antenna configurations. In the reflective room, however, the total suppression is much less, leading to the following result. 
\fi


	\ifIncludeMechanismDiscussion
		\subsection{Total Passive Suppression}
\begin{result}
\label{result:noAnalog}
In a low-reflection environment, combining directional isolation, absorptive shielding, and cross-polarization achieves sufficient passive suppression to eliminate the need for analog-domain self-interference cancellation at WiFi ranges. \note{(Not sure if I want this here, but we will see.)}
\end{result}


\begin{proof}[Reason for Result~\ref{result:noAnalog}] 
Existing full-duplex designs \cite{Bliss07SimultTX_RX, Khandani2010FDPatent, Radunovic:2009aa, Duarte10FullDuplex, Choi10FullDuplex, Jain2011RealTimeFD, Duarte11FullDuplex, Sahai11FullDuplex,Duarte2012FullDuplexWiFi, Duarte12Thesis} rely on a combination of both analog and adigital self-interference cancellation to provide the total suppression required for full-duplex to out-perform half-duplex. These designs must employ analog cancelers because, without significant passive suppression, the self-interference is 30-60~dB \note{(check)} more powerful that the signal-of-interest at typical WiFi ranges \note{\emph{[citation]}}. When the received signal is quantized, this large power differential causes the self-interference to dominate the dynamic range of the analog-to-digital converter (ADC), and the signal-of-interest loses 5-10 bits of precision (corresponding to the 30-60~dB differential). Even if the digital canceler perfectly nulls the self-interference, the signal-of-interest will suffer from prohibitive quantization noise.  The analysis in \cite{SahaiPhaseNoiseDraft} shows that such a concatenation of an analog canceler with a digital canceler provides no more total suppression than a digital canceler alone. The gain of using analog canceler is only to reduce quantization noise of the signal-of-interest. 

We see in Figure~\ref{fig:suppressionChamber} that, when in the low-reflection environment of the anechoic chamber, leveraging directional isolation, cross-polarization, absorptive shielding, in tandem ($\mathsf{DCA}$) provides more than 72~dB of passive self-interference suppression in all three configurations. In \cite{Duarte11FullDuplex}, 78~dB of \emph{total} self-interference suppression (both passive \emph{and} active) is achieved, which is shown to be sufficient for outperforming half-duplex for short-range (6.5~m) communication. By applying the three passive suppression, we are approaching the levels even before active cancellation is applied. Moreover, experimental characterizations have shown \note{\emph{[citation]}} that the path loss encountered in typical WiFi deployments ranges from 60~dB to 80~dB. If the self-interference is passively suppressed by $>70$~dB as seen in Figure~\ref{fig:suppressionChamber}, then self-interference will be at worst case 10~dB more powerful than the self-interference, which is negligible in terms of quantization noise of the signal-of-interest (1-2 bits lost). 
\end{proof}

\emph{Discussion of Design Implications of Result~\ref{result:noAnalog}:}
Implementation of an analog cancellation circuit requires significant design overhead and high-cost precision components. Digital cancellation, in contrast, can be simply incorporated in the DSP logic of the physical layer. For devices deployed in low-reflection environments, such as outdoor access~points, a deployment that leverages directional isolation, cross-polarization, and absorptive shielding for passive self-interference suppression can employ digital-only canceler to cancel the residual self-interference left behind by the passive suppression. \note{(As it stands, this is probably too weak to include. Need to talk to folks with more industry experience to get a better argument together.)} 

	\fi

	\subsection{Effect of Environmental Reflections}
\label{sec:reflections}


\begin{result}
\label{result:reflections}
	Environmental reflections limit the amount of passive self-interference suppression that can be achieved. 
\end{result}

Comparing Figures~\ref{fig:suppressionChamber}~and~\ref{fig:suppressionRoom}, we see that in every case the achieved suppression is less in the reflective room than in the anechoic chamber. Moreover, we observe that the incremental contribution of each applied mechanisms is much less in the reflective room than in the anechoic chamber. In Configuration~\dirBigA, for example, adding absorber improves the achieved suppression by 10~dB in the anechoic chamber, but provides less than 3~dB added suppression in the reflective room. 

\begin{proof}[Reason for Result~\ref{result:reflections}] 
Environmental reflections cause a bottleneck in passive self-interference suppression, because although the proposed mechanisms are very effective in suppressing the \emph{direct path} between the transmit and receive antennas, they are not necessarily effective in suppressing \emph{reflected} self-interference paths. Once the direct-path self-interference is suppressed to the extent that it is no longer the dominant path, further efforts to passively suppress self-interference may prove ineffective. 
Directional isolation is effective in suppressing the direct path, but will not suppress a reflected path that exits within the beamwidth of the transmit antenna and enters within the beamwidth of the receive antenna. 
Similarly, absorptive shielding certainly suppresses the direct path, but will do nothing to suppress a self-interference path that includes a reflector outside the node. 
Finally, reflections do not, in general, preserve polarization. Thus cross-polarizing the transmit and receive antennas suppresses the direct path, but not necessarily the reflected paths.

We will now observe the time-domain response of the measured self-interference channels to verify the hypothesis that suppression of the direct path but not reflected paths is the reason for Result~\ref{result:reflections}. 
We expect that for a given combination of passive suppression mechanisms the time response of the self-interference channel measured in the anechoic chamber will be nearly the same as that measured in the reflective room for an initial interval corresponding to the direct path delay, but afterward we expect the tail of the time response to be much more powerful in the reflective room than in the anechoic chamber. Moreover, we expect that when more passive suppression mechanisms are applied, the tail of the time response will become dominant over the direct path, explaining why adding more passive suppression mechanisms often does not improve the achieved suppression in the reflective room.  

%
\ifGenerateFigures
\begin{figure*}[htbp]
\renewcommand{\subcapsize}{\small}
\renewcommand{\subfigtopskip}{0pt}
\begin{center}
\subfigure[Omnis ($\emptyset$)\label{fig:OmniPDP}]{
	{\small
\begin{tikzpicture}
\begin{axis}[%
scale only axis,
width=0.30\textwidth,
height=0.21\textwidth,
xmin=0, xmax=100,
ymin=-110, ymax=-20,
xlabel={$t$ (ns)},
ylabel={$|h(t)|^2$ (dB)},
xlabel near ticks,
ylabel style={},
axis on top,
legend entries={${\sf \emptyset}$: Anechoic Chamber, ${\sf \emptyset}$: Reflective Room},
legend style={at={(0.5,1.0)},anchor=center,nodes=right}]
\addplot [color=KeynoteBlue,  line width=0.5pt]
table [x = time, y = O-Chamber]{./DocData/TimeDomainComparison.dat};
\addplot [color=KeynoteRed,  line width=0.5pt, densely dashed]
table [x = time, y = O-Room]{./DocData/TimeDomainComparison.dat};
\node[pin=60:{Reflective}] at (axis description cs:0.58,0.60) {};
\node[pin=60:{Anechoic}] at (axis description cs:0.59,0.20) {};
\end{axis}
\end{tikzpicture}

}
}
\subfigure[Directionals (${\sf D}$)\label{fig:DPDP}]{
	\hspace{-15pt}
	{\small
\begin{tikzpicture}
\begin{axis}[%
scale only axis,
width=0.30\textwidth,
height=0.21\textwidth,
xmin=0, xmax=100,
ymin=-110, ymax=-20,
xlabel={$t$ (ns)},
ylabel={$|h(t)|^2$ (dB)},
xlabel near ticks,
ylabel style={},
axis on top,
legend entries={${\sf D}$: Anechoic Chamber, ${\sf D}$: Reflective Room},
legend style={at={(0.5,1.0)},anchor=center,nodes=right}]
\addplot [color=KeynoteBlue,  line width=0.5pt]
table [x = time, y = D-Chamber]{./DocData/TimeDomainComparison.dat};
\addplot [color=KeynoteRed,  line width=0.5pt, densely dashed]
table [x = time, y = D-Room]{./DocData/TimeDomainComparison.dat};
\node[pin=60:{Reflective}] at (axis description cs:0.54,0.60) {};
\node[pin=60:{Anechoic}] at (axis description cs:0.38,0.16) {};
\end{axis}
\end{tikzpicture}

}

}
\subfigure[Directionals + Absorber (${\sf DA}$)\label{fig:DAPDP}]{
	{\small
\begin{tikzpicture}
\begin{axis}[%
scale only axis,
width=0.30\textwidth,
height=0.21\textwidth,
xmin=0, xmax=100,
ymin=-110, ymax=-20,
xlabel={$t$ (ns)},
ylabel={$|h(t)|^2$ (dB)},
xlabel near ticks,
ylabel style={},
axis on top,
legend entries={${\sf DA}$: Anechoic Chamber, ${\sf DA}$: Reflective Room},
legend style={at={(0.5,1.0)},anchor=center,nodes=right}]
\addplot [color=KeynoteBlue,  line width=0.5pt]
table [x = time, y = DA-Chamber]{./DocData/TimeDomainComparison.dat};
\addplot [color=KeynoteRed,  line width=0.5pt, densely dashed]
table [x = time, y = DA-Room]{./DocData/TimeDomainComparison.dat};
\node[pin=60:{Reflective}] at (axis description cs:0.50,0.57) {};
\node[pin=60:{Anechoic}] at (axis description cs:0.45,0.15) {};
\end{axis}
\end{tikzpicture}

}
}
\subfigure[Directionals + Crosspol + Absorber (${\sf DCA}$)\label{fig:DCAPDP}]{
	\hspace{-15pt}
	{\small
\begin{tikzpicture}
\begin{axis}[%
scale only axis,
width=0.30\textwidth,
height=0.21\textwidth,
xmin=0, xmax=100,
ymin=-110, ymax=-20,
xlabel={$t$ (ns)},
ylabel={$|h(t)|^2$ (dB)},
xlabel near ticks,
ylabel style={},
axis on top,
legend entries={${\sf DCA}$: Anechoic Chamber, ${\sf DCA}$: Reflective Room},
legend style={at={(0.5,1.0)},anchor=center,nodes=right}]
\addplot [color=KeynoteBlue,  line width=0.5pt]
table [x = time, y = DCA-Chamber]{./DocData/TimeDomainComparison.dat};
\addplot [color=KeynoteRed,  line width=0.5pt, densely dashed]
table [x = time, y = DCA-Room]{./DocData/TimeDomainComparison.dat};
\node[pin=60:{Reflective}] at (axis description cs:0.35,0.55) {};
\node[pin=60:{Anechoic}] at (axis description cs:0.25,0.06) {};
\end{axis}
\end{tikzpicture}

}
}
\caption{Comparing time responses measured in an anechoic chamber against the time responses measured in a reflective room. 
	\label{fig:EnvCmpr}}
\end{center}
\end{figure*}
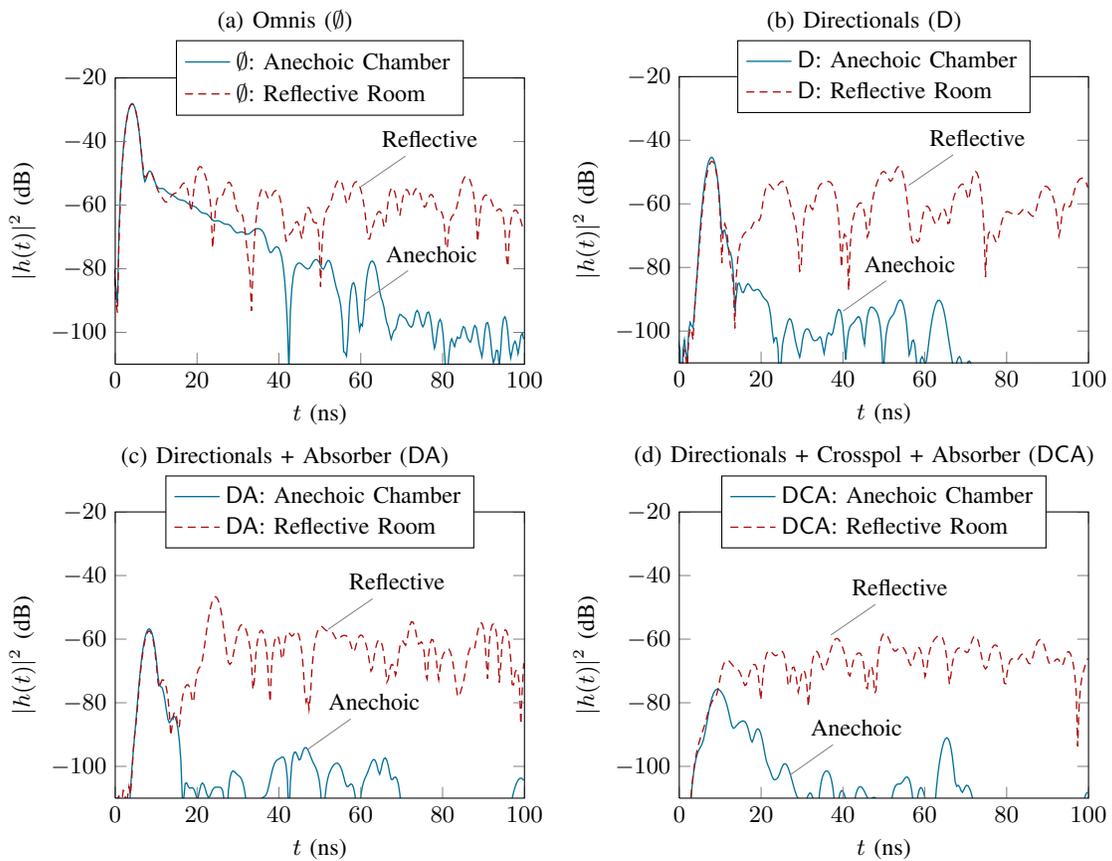
\fi

The hypothesized behavior is exactly what we observe in Figure~\ref{fig:EnvCmpr}, which compares the time domain channel responses measured in the chamber to those measured in the reflective room for four different combinations of passive suppression mechanisms. The time responses for omnidirectional antennas with no suppression mechanisms in place (Configuration~\omniBig, ${\sf mech=\emptyset}$) are shown in Figure~\ref{fig:OmniPDP}. 
In this case, the self-interference is dominated by the direct path in both environments, explaining why the value the average suppression, $\avgAlpha$, listed in Table~\ref{tab:Suppression} for this configuration is the same in both environments. 
Figure~\ref{fig:DPDP} shows that when we apply directional isolation (Configuration~\dirBigA, ${\sf mech=D}$), the direct path is significantly suppressed, but the reflection floor is not. We saw in Figure~\ref{fig:Suppression} that adding absorber between the directional antennas gives 10~dB improvement in the anechoic chamber, but less than 3~dB improvement in the reflective room; Figure~\ref{fig:DAPDP} explains why. Comparing Figure~\ref{fig:DPDP} and Figure~\ref{fig:DAPDP}, we see that adding absorber suppresses the \emph{direct path} by more than 10~dB in \emph{both} environments, but in the reflective environment the direct path is no longer the dominant path after adding absorber, while in the anechoic chamber the direct path remains the dominant path even after adding the absorber. Thus the full benefit of adding the absorber is not realized in the reflective environment. Finally, we see in Figure~\ref{fig:DCAPDP} that adding cross-polarization in addition to absorptive shielding adds another 10~dB of suppression in the anechoic chamber but does little in the reflective room, because reflections have already become the bottleneck. 
\end{proof}

\emph{Design Implications of Result~\ref{result:reflections}:}
All other factors being equal, a full-duplex infrastructure node should be deployed as far as possible from potential reflectors. In an environment that is unavoidably reflective, resources should not be expended attempting to passively suppress self-interference beyond the reflection strength. 
\ifLongDesignImplication
When an infrastructure node is deployed outdoors on a tower, it it may be worth the cost to employ all three passive suppression mechanisms.
In contrast, when deploying an access-point indoors, only one or two of the mechanisms may be sufficient to push the direct path below the reflection floor.
If an infrastructure node is deployed near reflectors that are \emph{static} and \emph{known}, then the proposed mechanisms may also be used to suppress reflected self-interference.
\fi

	\subsection{Impact of Passive Suppression Frequency Selectivity of Self-interference Channel}
\label{sec:coherence}

\begin{result}
\label{result:coherence}   
Passive self-interference suppression generally decreases the coherence bandwidth (i.e. increases the frequency selectivity) of the residual self-interference channel.
\end{result}

\ShowCBWTablesfalse
%
\ifGenerateFigures
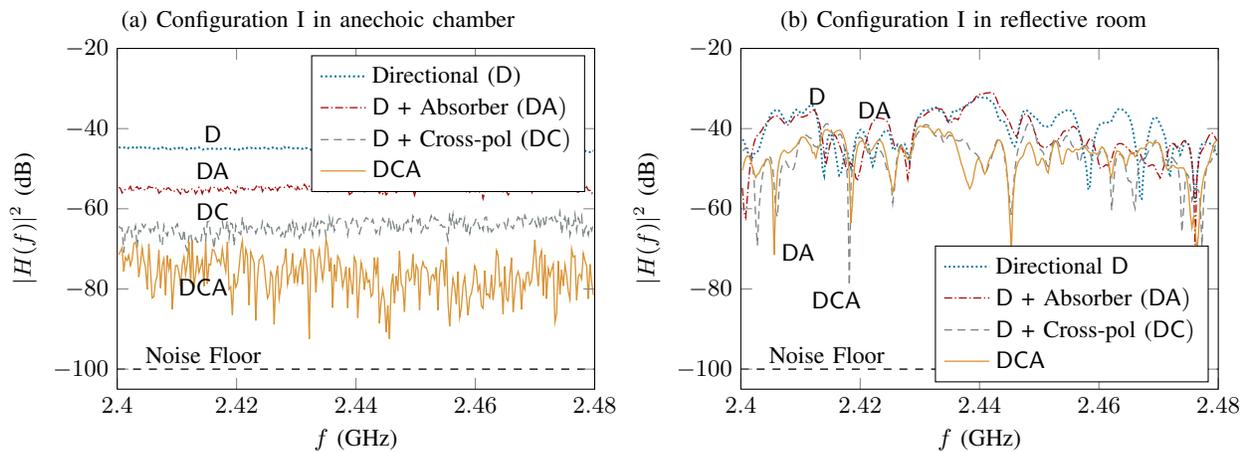
\begin{figure*}
\renewcommand{\subcapsize}{\small}
\renewcommand{\subfigcapskip}{-3pt}
\renewcommand{\subfigtopskip}{0pt}
	\centering
	\subfigure[Configuration~\dirBigA\ in anechoic chamber \label{fig:responseDirChamber}]{
		{\small
\begin{tikzpicture}
\begin{axis}[%
scale only axis,
width=0.35\textwidth,
height=0.25\textwidth,
xmin=2.4, xmax=2.48,
ymin=-105, ymax=-20,
xlabel={$f$ (GHz)},
ylabel={$|H(f)|^2$ (dB)},
xlabel near ticks,
ylabel near ticks,
ylabel style={},
axis on top,
legend entries={
Directional (${\sf D}$), 
${\sf D}$ + Absorber (${\sf DA})$,
${\sf D}$ + Cross-pol (${\sf DC})$,
${\sf DCA}$,
},
legend style={at={(0.98,0.98)},anchor=north east,nodes=right}]
\addplot [color=KeynoteBlue,  line width=0.75pt, densely dotted]
table [x = freq_GHz, y = None]{./DocData/ChamberConfigA.dat};
\node at (axis description cs:0.20,0.75) {$\sf D$};
\addplot [color=KeynoteRed,  line width=0.5pt, densely dashdotted]
table [x = freq_GHz, y = Absorber]{./DocData/ChamberConfigA.dat};
\node at (axis description cs:0.20,0.64) {$\sf DA$};
\addplot [color=KeynoteGray,  line width=0.5pt, densely dashed]
table [x = freq_GHz, y = Crosspol]{./DocData/ChamberConfigA.dat};
\node at (axis description cs:0.20,0.52) {$\sf DC$};
\addplot [color=KeynoteYellow,  line width=0.5pt]
table [x = freq_GHz, y = Crosspol+Absorber]{./DocData/ChamberConfigA.dat};
\node at (axis description cs:0.18,0.3) {$\sf DCA$};
\addplot [color=black,  line width=0.5pt, dashed]
coordinates{(2.4,-100) (2.48,-100)};
\node at (axis description cs:0.18,0.10) {Noise Floor};
\end{axis}
\ifShowCBWTables
	\node at (8.5,2.5)
	{\small \begin{tabular}{|c|c|c|}
			\hline
			${\sf mech}$ & $\avgAlpha$ & $B_C$ \tabularnewline
			\hline \hline
			${\sf D}$ & 45.2~dB & 19.59~MHz \tabularnewline \hline
			${\sf DA}$ & 55.2~dB & 12.58~MHz \tabularnewline \hline
			${\sf DC}$ & 64.3~dB & 3.94~MHz \tabularnewline \hline
			${\sf DCA}$ & 74.5~dB & 2.41~MHz \tabularnewline \hline		
		\end{tabular}};
\fi
\end{tikzpicture}
} 
	}
\ifLongFreSelectIllustration
	\subfigure[Configuration~\omniBig\ in anechoic chamber \label{fig:responseOmniChamber}]{
		\hspace{-20pt}
		{\small
\begin{tikzpicture}
\begin{axis}[%
scale only axis,
width=0.35\textwidth,
height=0.25\textwidth,
xmin=2.4, xmax=2.48,
ymin=-105, ymax=-20,
xlabel={$f$ (GHz)},
ylabel={$|H(f)|^2$ (dB)},
xlabel near ticks,
ylabel near ticks,
ylabel style={},
axis on top,
legend entries={
Omni (${\sf \emptyset}$), 
Absorber (${\sf A})$,
},
legend style={at={(0.98,0.02)},anchor=south east,nodes=right}]
\addplot [color=KeynoteBlue,  line width=0.7pt, densely dotted]
table [x = freq_GHz, y = None]{./DocData/ChamberConfigD.dat};
\node at (axis description cs:0.18,0.95) {$\sf \emptyset$};
\addplot [color=KeynoteRed,  line width=0.5pt, densely dashdotted]
table [x = freq_GHz, y = Absorber]{./DocData/ChamberConfigD.dat};
\node at (axis description cs:0.18,0.70) {$\sf A$};
\addplot [color=black,  line width=0.5pt, dashed]
coordinates{(2.4,-100) (2.48,-100)};
\node at (axis description cs:0.18,0.10) {Noise Floor};
\end{axis}
\ifShowCBWTables
	\node at (8.5,2.5)
	{\small \begin{tabular}{|c|c|c|}
			\hline
			${\sf mech}$ & $\avgAlpha$ & $B_C$ \tabularnewline
			\hline \hline
			${\sf \emptyset}$ & 27.8~dB  & 13.33~MHz \tabularnewline \hline
			${\sf A}$ & 48.0~dB  & 4.75~MHz \tabularnewline \hline		
		\end{tabular}};
\fi
\end{tikzpicture}
} \tabularnewline
	}
	\\
	\subfigure[Configuration~\omniBig\ in reflective room \label{fig:responseOmniRoom}]{
		{\small
\begin{tikzpicture}
\begin{axis}[%
scale only axis,
width=0.35\textwidth,
height=0.25\textwidth,
xmin=2.4, xmax=2.48,
ymin=-105, ymax=-20,
xlabel={$f$ (GHz)},
ylabel={$|H(f)|^2$ (dB)},
xlabel near ticks,
ylabel near ticks,
ylabel style={},
axis on top,
legend entries={
Omni (${\sf \emptyset}$), 
Absorber (${\sf A})$,
},
legend style={at={(0.98,0.02)},anchor=south east,nodes=right}]
\addplot [color=KeynoteBlue,  line width=0.7pt, densely dotted]
table [x = freq_GHz, y = None]{./DocData/RoomConfigD.dat};
\node at (axis description cs:0.16,0.96) {$\sf \emptyset$};
\addplot [color=KeynoteRed,  line width=0.5pt, densely dashdotted]
table [x = freq_GHz, y = Absorber]{./DocData/RoomConfigD.dat};
\node at (axis description cs:0.18,0.78) {$\sf A$};
\addplot [color=black,  line width=0.5pt, dashed]
coordinates{(2.4,-100) (2.48,-100)};
\node at (axis description cs:0.18,0.10) {Noise Floor};
\end{axis}
\ifShowCBWTables
	\node at (8.5,2.5)
	{\small \begin{tabular}{|c|c|c|}
			\hline
			${\sf mech}$ & $\avgAlpha$ & $B_C$ \tabularnewline
			\hline \hline
			${\sf \emptyset}$  & 27.8~dB  & 1.75~MHz \tabularnewline \hline
			${\sf A}$ & 40.1~dB & 0.74~MHz \tabularnewline \hline		
		\end{tabular}};
\fi
\end{tikzpicture}
}
	}
\fi	
	\subfigure[Configuration~\dirBigA\ in reflective room \label{fig:responseDirRoom}]{
		\hspace{-20pt}
		{\small
\begin{tikzpicture}
\begin{axis}[%
scale only axis,
width=0.35\textwidth,
height=0.25\textwidth,
xmin=2.4, xmax=2.48,
ymin=-105, ymax=-20,
xlabel={$f$ (GHz)},
ylabel={$|H(f)|^2$ (dB)},
xlabel near ticks,
ylabel near ticks,
ylabel style={},
axis on top,
legend entries={
Directional ${\sf D}$, 
${\sf D}$ + Absorber (${\sf DA})$,
${\sf D}$ + Cross-pol (${\sf DC})$,
${\sf DCA}$,
},
legend style={at={(0.98,0.02)},anchor=south east,nodes=right}]
\addplot [color=KeynoteBlue,  line width=0.7pt, densely dotted]
table [x = freq_GHz, y = None]{./DocData/RoomConfigA.dat};
\node at (axis description cs:0.16,0.86) {$\sf D$};
\addplot [color=KeynoteRed,  line width=0.5pt, densely dashdotted]
table [x = freq_GHz, y = Absorber]{./DocData/RoomConfigA.dat};
\node at (axis description cs:0.28,0.82) {$\sf DA$};
\addplot [color=KeynoteGray,  line width=0.5pt, densely dashed]
table [x = freq_GHz, y = Crosspol]{./DocData/RoomConfigA.dat};
\node at (axis description cs:0.12,0.4) {$\sf DA$};
\addplot [color=KeynoteYellow,  line width=0.5pt]
table [x = freq_GHz, y = Crosspol+Absorber]{./DocData/RoomConfigA.dat};
\node at (axis description cs:0.20,0.26) {$\sf DCA$};
\addplot [color=black,  line width=0.5pt, dashed]
coordinates{(2.4,-100) (2.48,-100)};
\node at (axis description cs:0.18,0.10) {Noise Floor};
%
\end{axis}
\ifShowCBWTables
	\node at (8.5,2.5)
	{\small \begin{tabular}{|c|c|c|}
		\hline
			${\sf mech}$ & $\avgAlpha$ & $B_C$ \tabularnewline
			\hline \hline
			${\sf D}$ & 38.1~dB & 0.71~MHz \tabularnewline \hline
			${\sf DA}$ & 38.8~dB & 0.82~MHz \tabularnewline \hline
			${\sf DC}$ & 44.7~dB & 0.83~MHz \tabularnewline \hline
			${\sf DCA}$ & 45.0~dB & 0.94~MHz \tabularnewline \hline		
		\end{tabular}};
\fi
\end{tikzpicture}
}
	}	
	\caption{Self-interference channel responses illustrating increased frequency selectivity with increased passive suppression. 
\label{fig:response}}
\end{figure*}
\fi

Figure~\ref{fig:response} shows measurements of the self-interference channel frequency response, $H^{(\mathsf{mech})}(f)$, that illustrate the trend of increased frequency selectivity with increased passive suppression.  Figure~\ref{fig:responseDirChamber} plots the channel responses measured for Configuration~\dirBigA\ in the anechoic chamber. We see in Figure~\ref{fig:responseDirChamber} that the variation of the self-interference channel gain with frequency increases as passive suppression mechanisms are applied.\footnote{The channel responses in Figure~\ref{fig:responseDirChamber} also may appear to be growing more noisy as more suppression mechanisms are applied. The appearance of greater noise is an artifact of the dB scaling of the plot making the measurement noise of the much lower-power suppressed channel appear larger. \revision{The thermal noise floor in this measurement was -100~dB, as computed from the manufacturer data sheet \cite{PNA_E8364B} given the 20Khz IFBW and averaging factor or 20 used in the measurement.}} 
\ifLongFreSelectIllustration
\else
The same trend was observed in the other four configurations measured in the anechoic chamber.
\fi
\ifLongFreSelectIllustration
The same trend was observed in the other directional-antenna configurations in the anechoic chamber.
Figure~\ref{fig:responseOmniChamber} shows the self-interference channel measurements for the omnidirectional antennas of Configuration~\omniBig\ in the anechoic chamber. Here we see that the self-interference channel becomes slightly more frequency selective when absorber is placed between antennas. The difference is much more pronounced in the reflective room, however, as shown in Figure~\ref{fig:responseOmniRoom}. 
\fi
Figure~\ref{fig:responseDirRoom} shows the channel responses measured for configuration~\dirBigA\ in the reflective room. We see in Figure~\ref{fig:responseDirRoom} that with directional isolation alone, the self-interference channel is already quite frequency selective. Adding cross-polarization and absorptive shielding in Figure~\ref{fig:responseDirRoom} does not significantly increase the frequency selectivity, but nor does it increase the average self-interference suppression, since the direct path has already been suppressed to the reflection floor as discussed in Result~\ref{result:reflections}. Figure~\ref{fig:responseDirRoom} hints that there is nothing special about directional isolation, cross-polarization, or absorptive shielding that causes the increase in frequency~selectivity, but the fact that each mechanism serves to reduce the \emph{direct path}.

\begin{proof}[Reason for Result~\ref{result:coherence}]
Result~\ref{result:coherence} is nearly a corollary of Result~\ref{result:reflections}. Since the dominant direct path is suppressed, the residual channel after passive suppression is the superposition of many reflected paths.
Passive suppression transforms the self-interference channel from a high-power line-of-sight-dominated channel to a low-power multi-path channel. It is well known that multi-path channels are frequency selective due to paths combining constructively at some frequencies and destructively at others. 
We will more precisely give the reason for Result~\ref{result:coherence} in two steps: first we will 
consider a simple analytical example that illustrates why the coherence bandwidth of the self-interference channel decreases as the direct path is suppressed. Second, we will present measurements of the coherence bandwidth for all five configurations in both environments to give further evidence of Result~\ref{result:coherence} and reinforce the intuitions gained from the analytical example.


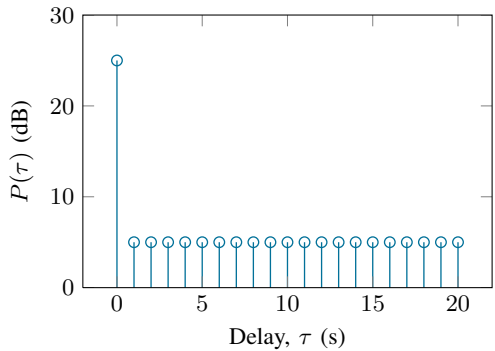
\begin{figure}[ht]
	\center
	{
\small
\begin{tikzpicture}
\begin{axis}[%
scale only axis,
width=0.30\textwidth,
height=0.20\textwidth,
ymin=0, ymax=30,
xlabel={Delay, $\tau$ (s)},
ylabel={$P(\tau)$ (dB)},
ylabel near ticks,
xlabel near ticks,
ylabel style={},
axis on top,
]
\addplot+[ycomb] [color=KeynoteBlue, mark=o,  line width=0.5pt]
table [x = tau, y = PDP]{./DocData/ExamplePDP.dat};
\end{axis}
\end{tikzpicture}
}
	\caption{\revision{Example of a power-delay profile with both direct and reflected paths. In this example
	 $P_D = 25$~dB, $P_R = 5$~dB $\Rightarrow \DRR= 20$~dB, and $T_R = 20$~s.}
	\label{fig:NTapPDP}}
\end{figure}

\subsubsection{Analytical example}
\revision{

Consider the ``two-level'' self-interference power delay profile shown in Figure~\ref{fig:NTapPDP}, in which the first tap which contains the direct path and has power $P_D$, and  the remaining $T_R-1$ taps, all due to reflected paths, have power $P_R$, where $T_R$ is the reflection duration of the environment. 
}
Denote the ratio of the direct-path power to the reflected-path power as the direct-to-reflected ratio,
$\DRR \equiv \frac{P_D}{P_R}$. 
%
As passive suppression mechanisms are employed, we know from Result~\ref{result:reflections} that $\DRR$ will decrease, since the direct path will be suppressed while the reflective paths maintain the same power. 



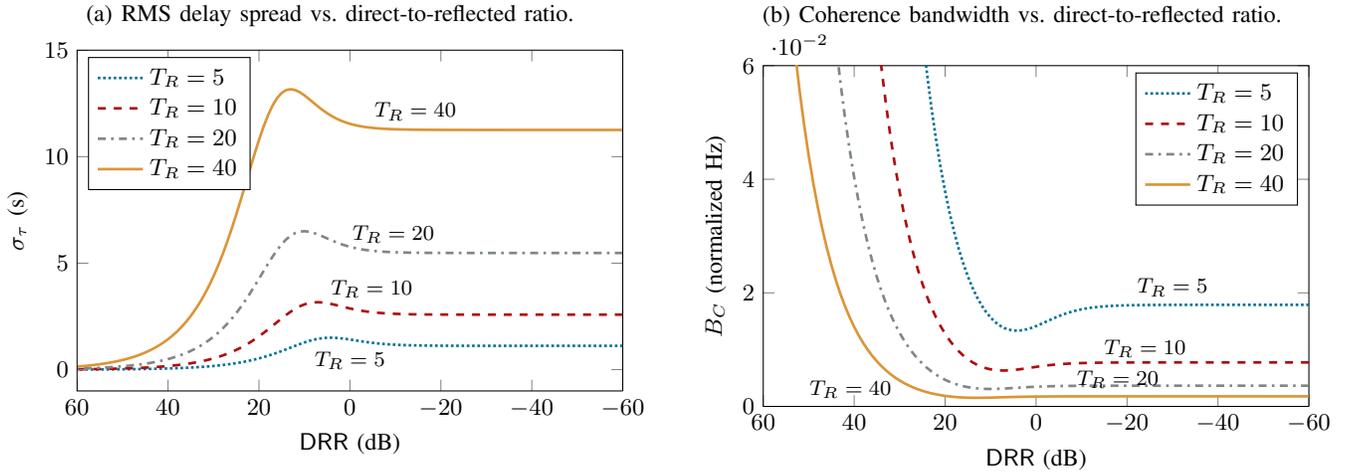
\begin{figure*}[ht]
	\renewcommand{\subcapsize}{\small}
	\renewcommand{\subfigtopskip}{-3pt}
	\ifCLASSOPTIONonecolumn
	\renewcommand{\subfigtopskip}{-10pt}
	\fi
	\center
	\subfigure[RMS delay spread vs. direct-to-reflected ratio.]{
		{
\small
\begin{tikzpicture}
\begin{axis}[%
scale only axis,
width=0.40\textwidth,
height=0.25\textwidth,
xmin=-60, xmax = 60,
ymin=-1, ymax=15,
xlabel={$\mathsf{DRR}$ (dB)},
ylabel={$\sigma_{\tau}$ (s)},
x dir = reverse,
ylabel near ticks,
xlabel near ticks,
ylabel style={},
axis on top,
legend entries={
$T_R = 5$,
$T_R = 10$, 
$T_R = 20$,
$T_R = 40$
},
legend style={at={(0.02,0.98)},anchor=north west,nodes=right}]
\addplot [color=KeynoteBlue,  line width=1.00pt, densely dotted]
table [x = DRR, y = Beta0]{./DocData/RMSDelaySpread.dat};
\node at (axis description cs:0.5,0.09) {\footnotesize$T_R = 5$};
\addplot [color=KeynoteRed,  line width=1.00pt, dashed]
table [x = DRR, y = Beta1]{./DocData/RMSDelaySpread.dat};
\node at (axis description cs:0.54,0.30) {\footnotesize$T_R = 10$};
\addplot [color=KeynoteGray,  line width=1.00pt, dashdotted]
table [x = DRR, y = Beta5]{./DocData/RMSDelaySpread.dat};
\node at (axis description cs:0.58,0.46) {\footnotesize$T_R = 20$};
\addplot [color=KeynoteYellow,  line width=1.00pt]
table [x = DRR, y = Beta10]{./DocData/RMSDelaySpread.dat};
\node at (axis description cs:0.62,0.82) {\footnotesize$T_R = 40$};
\end{axis}
\end{tikzpicture}
}
		\label{fig:RMSExample}
	}\subfigure[Coherence bandwidth vs. direct-to-reflected ratio.]{
		{
\small
\begin{tikzpicture}
\begin{axis}[%
scale only axis,
width=0.40\textwidth,
height=0.25\textwidth,
xmin=-60, xmax = 60,
ymin=0, ymax=0.06,
xlabel={$\mathsf{DRR}$ (dB)},
x dir = reverse,
ylabel={$B_{C}$ (normalized Hz)},
ylabel near ticks,
xlabel near ticks,
ylabel style={},
yticklabel=\pgfmathprintnumber{\tick},
scaled y ticks=base 10:2
axis on top,
legend entries={
$T_R = 5$,
$T_R = 10$, 
$T_R = 20$,
$T_R = 40$
},
legend style={at={(0.98,0.98)},anchor=north east,nodes=right}]
\addplot [color=KeynoteBlue,  line width=1.00pt, densely dotted]
table [x = DRR, y = Beta0]{./DocData/ExampleCBW.dat};
\node at (axis description cs:0.75,0.35) {\footnotesize$T_R = 5$};
\addplot [color=KeynoteRed,  line width=1.00pt, dashed]
table [x = DRR, y = Beta1]{./DocData/ExampleCBW.dat};
\node at (axis description cs:0.7,0.17) {\footnotesize$T_R = 10$};
\addplot [color=KeynoteGray,  line width=1.00pt, dashdotted]
table [x = DRR, y = Beta5]{./DocData/ExampleCBW.dat};
\node at (axis description cs:0.65,0.08) {\footnotesize$T_R = 20$};

\addplot [color=KeynoteYellow,  line width=1.00pt]
table [x = DRR, y = Beta10]{./DocData/ExampleCBW.dat};
\node at (axis description cs:0.16,0.05) {\footnotesize$T_R = 40$};
\end{axis}
\end{tikzpicture}
}
		\label{fig:CBWExample}
	}
	\caption{Impact passive suppression on RMS delay spread and coherence bandwidth.
	\label{fig:NtapExample}}
\end{figure*}

Figure~\ref{fig:NtapExample} plots the RMS delay spread, $\sigma_{\tau}$, and the corresponding coherence bandwidth, $B_C = 0.02/\sigma_\tau$, vs. $\DRR$ for four different values of $T_R$. 
Suppressing the direct path (reducing $\DRR$) will lead to a heavier tail in the normalized power delay profile, $P_0(\tau)$, and hence a larger RMS delay spread and smaller coherence bandwidth. When the passive suppression is such that the power in the direct path is the same as the power in the reflected paths, the RMS delay spread is maximum. Once $\DRR < 0$~dB, the reflected portion of the power delay profile becomes dominant, so that further suppression of the direct path has little impact on the RMS delay spread and coherence bandwidth. 

In summary, the analytical example shows the following trend. When no passive suppression is applied, the direct-path self-interference will be significantly stronger than reflected paths, and the coherence bandwidth will be relatively high, such that the self-interference channel will be relatively frequency-flat. As passive suppression is applied, the reflected paths become a larger portion of the power delay profile, and the coherence bandwidth decreases such that the self-interference channel becomes more frequency-selective. In the limit of very strong passive suppression, the direct path is effectively eliminated, and the coherence bandwidth levels off, because the delay spread converges to the reverberation spread of the environment.

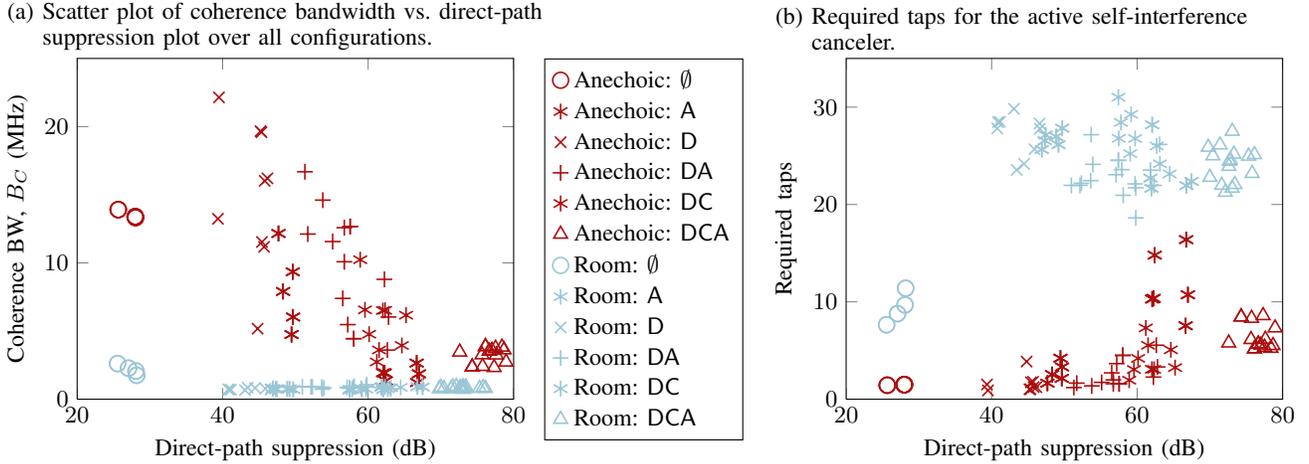
\begin{figure*}[t]
\renewcommand{\subcapsize}{\small}
\renewcommand{\subfigcapmargin}{10pt}
\subfigure[Scatter plot of coherence bandwidth vs. direct-path \newline suppression plot over all configurations.\label{fig:MechanismScatter}]{
{
\small
\begin{tikzpicture}
\begin{axis}[%
scale only axis,
width=0.32\textwidth,
height=0.25\textwidth,
xmin=20, xmax=80,
ymin=0, ymax=25,
xlabel={Direct-path suppression (dB)},
ylabel={Coherence BW, $B_{C}$ (MHz)},
ylabel near ticks,
xlabel near ticks,
ylabel style={},
axis on top,
mark size=3,
every axis plot post/.append style={semithick},
legend entries={
Anechoic: ${\sf \emptyset}$,
Anechoic: ${\sf A}$,
Anechoic: ${\sf D}$,
Anechoic: ${\sf DA}$,
Anechoic: ${\sf DC}$,
Anechoic: ${\sf DCA}$,
Room: ${\sf \emptyset}$,
Room: ${\sf A}$,
Room: ${\sf D}$,
Room: ${\sf DA}$,
Room: ${\sf DC}$,
Room: ${\sf DCA}$,
},
legend style={at={(1.07,1.0)},anchor=north west,nodes=right}
]
\addplot+[only marks] [color=KeynoteRed, mark=o]
table [x = directPathSuppression, y = CBW]{./DocData/ChamberO.dat};
\addplot+[only marks] [color=KeynoteRed,  mark=asterisk]
table [x = directPathSuppression, y = CBW]{./DocData/ChamberOA.dat};
%
\addplot+[only marks] [color=KeynoteRed,  mark=x]
table [x = directPathSuppression, y = CBW]{./DocData/ChamberD.dat};
\addplot+[only marks] [color=KeynoteRed,  mark=+]
table [x = directPathSuppression, y = CBW]{./DocData/ChamberDA.dat};
\addplot+[only marks] [color=KeynoteRed,  mark=asterisk]
table [x = directPathSuppression, y = CBW]{./DocData/ChamberDC.dat};
\addplot+[only marks] [color=KeynoteRed,  mark=triangle]
table [x = directPathSuppression, y = CBW]{./DocData/ChamberDCA.dat};
\addplot+[only marks] [color=KeynoteBlue!40, mark=o]
table [x = directPathSuppression, y = CBW]{./DocData/RoomO.dat};
\addplot+[only marks] [color=KeynoteBlue!40,  mark=asterisk]
table [x = directPathSuppression, y = CBW]{./DocData/RoomOA.dat};
\addplot+[only marks] [color=KeynoteBlue!40,  mark=x]
table [x = directPathSuppression, y = CBW]{./DocData/RoomD.dat};
\addplot+[only marks] [color=KeynoteBlue!40,  mark=+]
table [x = directPathSuppression, y = CBW]{./DocData/RoomDA.dat};
\addplot+[only marks] [color=KeynoteBlue!40,  mark=asterisk]
table [x = directPathSuppression, y = CBW]{./DocData/RoomDC.dat};
\addplot+[only marks] [color=KeynoteBlue!40,  mark=triangle]
table [x = directPathSuppression, y = CBW]{./DocData/RoomDCA.dat};
\end{axis}
\end{tikzpicture}
}
} \hspace{-15pt}
\subfigure[Required taps for the active self-interference canceler.
	\label{fig:taps}]{
	{
\small
\begin{tikzpicture}
\begin{axis}[%
scale only axis,
width=0.32\textwidth,
height=0.25\textwidth,
xmin=20, xmax=80,
ymin=0, ymax=35,
xlabel={Direct-path suppression (dB)},
ylabel={Required taps},
ylabel near ticks,
xlabel near ticks,
ylabel style={},
axis on top,
mark size=3,
every axis plot post/.append style={semithick},
legend style={at={(1.05,1.0)},anchor=north west,nodes=right}
]
\addplot+[only marks] [color=KeynoteRed, mark=o]
table [x = directPathSuppression, y = filterTaps]{./DocData/ChamberO.dat};
\addplot+[only marks] [color=KeynoteRed,  mark=asterisk]
table [x = directPathSuppression, y = filterTaps]{./DocData/ChamberOA.dat};
%
\addplot+[only marks] [color=KeynoteRed,  mark=x]
table [x = directPathSuppression, y = filterTaps]{./DocData/ChamberD.dat};
\addplot+[only marks] [color=KeynoteRed,  mark=+]
table [x = directPathSuppression, y = filterTaps]{./DocData/ChamberDA.dat};
\addplot+[only marks] [color=KeynoteRed,  mark=asterisk]
table [x = directPathSuppression, y = filterTaps]{./DocData/ChamberDC.dat};
\addplot+[only marks] [color=KeynoteRed,  mark=triangle]
table [x = directPathSuppression, y = filterTaps]{./DocData/ChamberDCA.dat};
\addplot+[only marks] [color=KeynoteBlue!40, mark=o]
table [x = directPathSuppression, y = filterTaps]{./DocData/RoomO.dat};
\addplot+[only marks] [color=KeynoteBlue!40,  mark=asterisk]
table [x = directPathSuppression, y = filterTaps]{./DocData/RoomOA.dat};
\addplot+[only marks] [color=KeynoteBlue!40,  mark=x]
table [x = directPathSuppression, y = filterTaps]{./DocData/RoomD.dat};
\addplot+[only marks] [color=KeynoteBlue!40,  mark=+]
table [x = directPathSuppression, y = filterTaps]{./DocData/RoomDA.dat};
\addplot+[only marks] [color=KeynoteBlue!40,  mark=asterisk]
table [x = directPathSuppression, y = filterTaps]{./DocData/RoomDC.dat};
\addplot+[only marks] [color=KeynoteBlue!40,  mark=triangle]
table [x = directPathSuppression, y = filterTaps]{./DocData/RoomDCA.dat};
\end{axis}
\end{tikzpicture}
}
}
\caption{Influence of direct-path suppression of frequency selectivity
\label{fig:scatter}}
\end{figure*}

\subsubsection{Coherence Bandwidth Measurements}
Now let us check that the measurements in the passive suppression characterization follow the trend illustrated in the above analytical example. For each self-interference channel measurement, the the direct-path suppression and the coherence bandwidth were computed from the measured frequency response. The resulting tuples of direct-path suppression and coherence bandwidth are shown as a scatter plot in Figure~\ref{fig:MechanismScatter}. 
The dark red points in Figure~\ref{fig:MechanismScatter} correspond to measurements made in the anechoic chamber, and the light blue points correspond to the reflective room. The combination of mechanisms used is labeled by the symbol as shown in the legend.\footnote{Not evident in Figure~\ref{fig:scatter} (to reduce clutter and focusing on the trend), is the configuration on which the measurement was made, which accounts for the sometimes wide variation in points corresponding to the same suppression mechanisms.}


%

The trend we observe in Figure~\ref{fig:MechanismScatter} is the same as predicted from our analytical example. The anechoic chamber has a very low reflection floor, around $-80$~dB, so when no suppression is applied, the coherence bandwidth is large: between 10 and 25 MHz depending on the configuration. If absorptive shielding or cross-polarization is employed: the direct-path suppression is increased to 50-70~dB and the coherence bandwidth shrinks to 2-15~MHz. When both cross-polarization and absorptive shielding are applied together, the direct-path suppression is greater than 70~dB, near the reflection floor of the chamber, hence the coherence bandwidth begins to level off at around 3~MHz.
%
In the reflective room, the trend is the same but much more abrupt, since the reflection floor is much higher: around $-45$~dB. In the reflective room, only when omnidirectional antennas are used with no suppression mechanisms ($\sf{mech = \emptyset}$) is the direct path significantly stronger than the reflected paths (as was seen in Figure~\ref{fig:EnvCmpr}), in which case the coherence bandwidth is around 2~MHz. Once one or more suppression mechanisms are employed, the direct path is pushed below 40~dB, and the coherence bandwidth converges to the 0.6-0.9~MHz dictated by the long reverberations of the reflective room.
Typical deployments will have a reflection floor between the two extremes presented here, but the trend will be the same. Passive suppression can transform a relatively frequency-flat self-interference channel to a highly frequency-selective channel, but the coherence bandwidth will be no worse than natural coherence bandwidth dictated by the reverberation depth of the environment. 

\end{proof}



\emph{Design Implications of Result~\ref{result:coherence}:}
Most full-duplex systems will attempt to further suppress self-interference by employing active cancellation in addition to passive suppression. Self-interference is cancelled by leveraging foreknowledge of the signal to inject a cancellation waveform into the received signal path. 
\revision{
To craft the cancellation signal, a copy of the transmit signal must pass through a filter whose response is the negative of the self-interference channel. Result~\ref{result:coherence} implies that the number of taps in this cancellation filter may increase when passive suppression is employed. Figure~\ref{fig:taps} takes the data shown in Figure~\ref{fig:MechanismScatter}, and computes the required number of taps in the cancellation filter, assuming the transmit signal has a bandwidth of 20~MHz. In the low-reflection environment only one or two taps are needed when the passive suppression is low (25-45~dB), but as the passive suppression gets strong (60-80~dB), 10-15 taps may be needed. In the high-reflection environment, 7-12 taps are needed when the passive suppression is  low, but once the passive suppression is more than 40~dB, nearly 30 taps are needed. 
} 
In an OFDM system, Result~\ref{result:coherence} implies that when active cancellation is employed in concatenation with strong passive suppression, it may be better to perform the active cancellation on a per-subcarrier basis as in \cite{Duarte12Thesis, Duarte2012FullDuplexWiFi} rather than directly canceling the broadband self-interference as in \cite{Jain2011RealTimeFD}.

\revision{

\subsection{Passive Suppression vs. Active Cancellation}


The above discussion raises the following question: is there a tradeoff between passive suppression and active cancellation, and would one rather use one than the other, or always use both together? 
The results from this paper and those from related works are in agreement that \emph{passive suppression and active cancellation should be applied together whenever possible.} 
In this paper, it is shown that passive suppression is limited by reflections, and active cancellation is needed to suppress the residual reflected paths. Conversely, in \cite{SahaiPhaseNoiseDraft,Day12FDRelay,Day12FDMIMO} it is shown that active cancellation is limited by transmitter noise, and thus passive suppression (which can suppress transmitter noise just as well as transmitter signal) is needed to improve performance. 
In \cite{SahaiPhaseNoiseDraft} a preprint of this paper is cited and our Result~\ref{result:reflections} is used to show that although passive suppression can degrade active cancelation performance, it can only increase the \emph{total} self-interference suppression. 
In the full-duplex system evaluation such as \cite{Duarte2012FullDuplexWiFi,Aryafar12MIDU,Khandani2010FDPatent}, the designs employing both passive suppression and active cancellation together always have the best performance. 
The summary of ours and the previous work is that although the gains of passive suppression and active cancellation are not dB-additive, 
never is it the case that passive suppression alone or active cancellation alone will have better performance than both in concatenation. 
The question of which approach, passive suppression or active cancellation, is more \emph{efficient} is still open, and is in interesting topic for future study. 
}

\revision{
\section{Impact of Passive Suppression on System Metrics}
\label{sec:analysis}
We will now analytically assess the impact that Results~1~and~2 have on the performance of a full-duplex infrastructure node.  
Consider a full-duplex infrastructure node that receives an uplink signal from one user while also transmitting to another user as shown in Figure~\ref{fig:non-recip}. We will focus on uplink performance, since it is the uplink that suffers from self-interference. 

\subsection{Channel Model}
The channel model must capture the limitations of both passive suppression and active cancellation. Prior work \cite{SahaiPhaseNoiseDraft,Day12FDMIMO,Day12FDRelay} has identified transmitter noise as the major factor limiting active cancellation, and this paper identifies environmental reflections as the major factor limiting passive suppression. Thus we model the signal received by the full-duplex infrastructure node at time $n$ as,
\ifCLASSOPTIONonecolumn
	\begin{equation}
	y[n] = h_S[n]*(x_S[n] + z_S[n]) + h_I[n]*(x_I[n] + z_I[n]) + z_R[n],
	\end{equation}
\else
	\begin{align}
	y[n] = &h_S[n]*(x_S[n] + z_S[n]) \\ \nonumber
	 &+ h_I[n]*(x_I[n] + z_I[n]) + z_R[n],\nonumber
	\end{align}
\fi
where $x_I$ and $x_S$ are the signals transmitted by the infrastructure node and the uplink user, respectively, $h_I$ and $h_S$ are the discrete baseband-equivalent impulse responses of the self-interference channel and the signal-of-interest channel, respectively, $z_I$ and $z_S$  are transmitter noise in the self-interference and signal-of-interest, respectively, $z_R$ is the receiver noise, and ``$*$'' signifies convolution. For convenience, we assume the signal-of-interest channel, $h_S$, has only a single tap and the signal-of-interest is free of transmitter noise, as these will have little impact on self-interference suppression. 
Let the receiver noise and transmitter noise be drawn i.i.d. from circularly-symmetric gaussian distributions of variance $N_R$ and $N_T$, respectively, and let the transmit signals, $x_S$ and $x_I$, be subject to average power constraint of $P_T$. 
With the above assumptions, the received signal can be rewritten as
\ifCLASSOPTIONonecolumn
	\begin{equation}
	\label{eq:FDChannel}
	y[n] =  h_S x_S[n] +  h_I[n]*(x_I[n] + z_I[n]) + z_R[n], \qquad 	z_I\sim \mathcal{CN}(0,N_T),\ z_R\sim \mathcal{CN}(0,N_R).
	\end{equation}
\else
	\begin{align}
	\label{eq:FDChannel}
	y[n] =  h_S x_S[n] +  h_I[n]*(x_I[n] + z_I[n]) + z_R[n], \nonumber \\ 	z_I\sim \mathcal{CN}(0,N_T),\ z_R\sim \mathcal{CN}(0,N_R). 
	\end{align}
\fi


Let us extend the ``two-level'' self-interference power delay profile discussed in section~\ref{sec:coherence} to model the self-interference channel $h_I[n]$. We assume that there are many reflected paths of power $P_R$ that extend over $T_R$ taps and a single direct path of power $P_D$.  Thus we model the first channel tap as Rician fading with $\kappa$-factor $P_D/P_R$ and variance $P_D + P_R$, and the remaining $T_R-1$ taps as Rayleigh fading with variance $P_R$. We assume the single-tap signal-of-interest channel, $h_S$, is Rayleigh fading with variance $P_S$. Per Result 1, passive suppression can reduce $P_D$, but not $P_R$. We assume each channel tap is ergodic block fading with coherence time $T_C \gg 1$, and assume coding can be performed over a large number of coherence times so that the notion of ergodic capacity can be employed \cite{TseBook}. We assume both the user and the infrastructure node have perfect knowledge of the channel responses, and the infrastructure has knowledge of its own transmit signal, $x_I$, but has no knowledge of its transmitter noise, $z_I$. 

\subsection{Impact of Direct-path Suppression on Capacity}

\begin{theorem}
\label{thm:ergCap}
The capacity of the full-duplex uplink channel described in eq.~\ref{eq:FDChannel} is
\begin{equation}
\label{eq:FDCap}	
C_{\rm Uplink}^{\rm FD} = \Ex \left[\int_{-\pi}^{\pi} \left( 1 + \frac{(\nu - S_N(f))^+}{S_N(f)} \right) d f \right],
\end{equation}
where $S_N(f) = |\sqrt{N_T}H_I(f) + \sqrt{N_R}|^2$, $H_I(f)$ is the DTFT of the self-interference channel impulse response, $h_I[n]$, and $\nu$ is a water-filling parameter chosen such that $\int(\nu - S_N(f))^+ df = | h_S |^2 P_T$. The expectation is with respect to the channel responses, $h_I$ and $h_S$.
\end{theorem}

\begin{IEEEproof}
Because the portion of the self-interference due to intended signal, $h_I*x_I$, is known, it can be cancelled by forming 
\ifCLASSOPTIONonecolumn
	\begin{equation}
	y'[n] = y[n] - h_I[n]*x_I[n]  
		= h_Sx_S[n] + h_I[n]*z_I[n] +  z_R[n] \label{eq:cancelled}
	\end{equation}
\else
	\begin{align}
	y'[n] &= y[n] - h_I[n]*x_I[n]  \\
	&= h_Sx_S[n] + h_I[n]*z_I[n] +  z_R[n] \label{eq:cancelled},
	\end{align}
\fi
so that the only self-interference that remains is the transmitter noise filtered by the self-interference channel response. 
%
%
Within a single coherence time, Equation~\ref{eq:cancelled} is equivalent to the well-known colored Gaussian noise channel with input power constraint $P = | h_S |^2 P_T$ and noise power spectral density $S_N(f) = |\sqrt{N_T}H_I(f) + \sqrt{N_R}|^2$, where $H_I(f)$ is the DTFT of $h_I[n]$.
The capacity of the colored Gaussian channel is achieved via water-filling in the spectral domain and is given by \cite{CoverThomas}
\begin{equation}
C = \int_{-\pi}^{\pi} \left( 1 + \frac{(\nu - S_N(f))^+}{S_N(f)} \right) d f,
\end{equation}
where $\nu$ is the ``water level'' chosen such that $\int(\nu - S_N(f))^+ df = P$. By averaging over many independent fades of the channel taps, the capacity given in Theorem~\ref{thm:ergCap} is obtained.   
\end{IEEEproof}
}

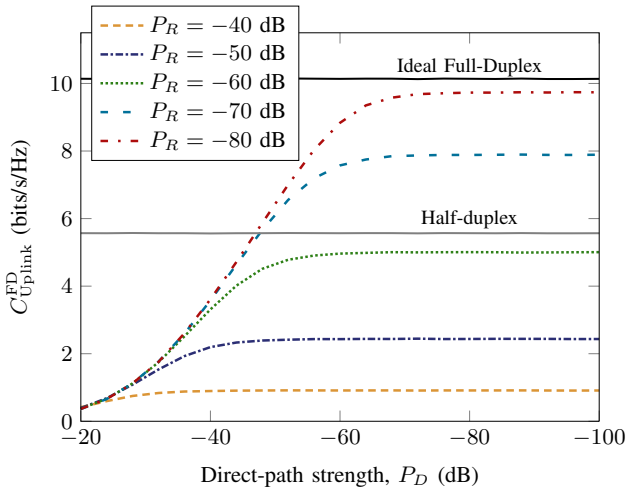
\begin{figure}[htdp]
\centering
	\renewcommand{\subcapsize}{\normalsize}
	\renewcommand{\subfigcapskip}{5pt}
\subfigure[System parameters assumed. \label{tab:simParams}]{
\renewcommand{\arraystretch}{1.0}
\centering
{\small
\begin{tabular}{r|l}
\emph{Parameter} & \emph{Value} \tabularnewline \hline
Transmit power, $P_T$ & $0$~dBm \tabularnewline 
Transmit noise power, $N_T$ & $-30$~dBm \tabularnewline 
Receiver noise power, $N_R$ & $-90$~dBm \tabularnewline 
Signal channel variance, $P_S$ & $-60$~dB \tabularnewline 
Extent of reflections, $T_R$ & 32 \tabularnewline 
\end{tabular}
}
}
\subfigure[Capacity vs. direct-path strength.\label{fig:CapPlot}]{
\centering
\ifCLASSOPTIONonecolumn
{
\small
\begin{tikzpicture}
\begin{axis}[%
scale only axis,
width=0.4\textwidth,
height=0.3\textwidth,
xmin=-100, xmax = -20,
ymin=0, 
ymax=11.5,
xlabel={Direct-path strength, $P_D$ (dB)},
ylabel={$C_{\rm Uplink}^{\rm FD}$ (bits/s/Hz)},
x dir = reverse,
ylabel near ticks,
ylabel style={},
axis on top,
legend entries={
$P_R = -40$~dB,
$P_R = -50$~dB, 
$P_R = -60$~dB,
$P_R = -70$~dB,
$P_R = -80$~dB,
},
legend style={at={(0.02,1.07)},anchor=north west,nodes=right}]
\addplot [color=KeynoteYellow,  line width=1.00pt, densely dashed]
table [x = directPathAtten_dB, y = FD40dB]{./DocData/RateVsSuppresion.dat};
\addplot [color=ECEpurple,  line width=1.00pt, densely dashdotted]
table [x = directPathAtten_dB, y = FD50dB]{./DocData/RateVsSuppresion.dat};
%
\addplot [color=KeynoteGreen,  line width=1.00pt, densely dotted]
table [x = directPathAtten_dB, y = FD60dB]{./DocData/RateVsSuppresion.dat};
\addplot [color=KeynoteBlue,  line width=1.00pt, loosely dashed]
table [x = directPathAtten_dB, y = FD70dB]{./DocData/RateVsSuppresion.dat};
%
\addplot [color=KeynoteRed,  line width=1.00pt, loosely dashdotted]
table [x = directPathAtten_dB, y = FD80dB]{./DocData/RateVsSuppresion.dat};
\addplot [color=gray,  line width=0.8pt ]
table [x = directPathAtten_dB, y = HD]{./DocData/RateVsSuppresion.dat};\node at (axis description cs:0.75,0.91) {\footnotesize Ideal Full-Duplex};
\addplot [color=black,  line width=0.8pt]
table [x = directPathAtten_dB, y = IdealFD]{./DocData/RateVsSuppresion.dat};
\node at (axis description cs:0.75,0.52) {\footnotesize Half-duplex};
\end{axis}
\end{tikzpicture}
}
\else
\scalebox{0.95}{{
\small
\begin{tikzpicture}
\begin{axis}[%
scale only axis,
width=0.4\textwidth,
height=0.3\textwidth,
xmin=-100, xmax = -20,
ymin=0, 
ymax=11.5,
xlabel={Direct-path strength, $P_D$ (dB)},
ylabel={$C_{\rm Uplink}^{\rm FD}$ (bits/s/Hz)},
x dir = reverse,
ylabel near ticks,
ylabel style={},
axis on top,
legend entries={
$P_R = -40$~dB,
$P_R = -50$~dB, 
$P_R = -60$~dB,
$P_R = -70$~dB,
$P_R = -80$~dB,
},
legend style={at={(0.02,1.07)},anchor=north west,nodes=right}]
\addplot [color=KeynoteYellow,  line width=1.00pt, densely dashed]
table [x = directPathAtten_dB, y = FD40dB]{./DocData/RateVsSuppresion.dat};
\addplot [color=ECEpurple,  line width=1.00pt, densely dashdotted]
table [x = directPathAtten_dB, y = FD50dB]{./DocData/RateVsSuppresion.dat};
%
\addplot [color=KeynoteGreen,  line width=1.00pt, densely dotted]
table [x = directPathAtten_dB, y = FD60dB]{./DocData/RateVsSuppresion.dat};
\addplot [color=KeynoteBlue,  line width=1.00pt, loosely dashed]
table [x = directPathAtten_dB, y = FD70dB]{./DocData/RateVsSuppresion.dat};
%
\addplot [color=KeynoteRed,  line width=1.00pt, loosely dashdotted]
table [x = directPathAtten_dB, y = FD80dB]{./DocData/RateVsSuppresion.dat};
\addplot [color=gray,  line width=0.8pt ]
table [x = directPathAtten_dB, y = HD]{./DocData/RateVsSuppresion.dat};\node at (axis description cs:0.75,0.91) {\footnotesize Ideal Full-Duplex};
\addplot [color=black,  line width=0.8pt]
table [x = directPathAtten_dB, y = IdealFD]{./DocData/RateVsSuppresion.dat};
\node at (axis description cs:0.75,0.52) {\footnotesize Half-duplex};
\end{axis}
\end{tikzpicture}
}}
\fi 
}
\caption{Numerical evaluation of ergodic uplink capacity of Eq.~\ref{eq:FDCap}.
	\label{fig:CapEval}}
\end{figure}

%

\revision{
We want to compare full-duplex capacity to both half-duplex and ``ideal'' full-duplex capacity.
Assuming that in half-duplex mode the uplink must evenly time-share with the downlink, the uplink capacity is easily shown to be
\begin{equation}
\label{eq:HDCap}
C_{\rm Uplink}^{\rm HD} = \Ex \left[\frac{1}{2} \log_2\left(1 + \frac{\ 2 | h_S |^2 P_T}{N_R} \right) \right],
\end{equation}
where the pre-log factor of $1/2$ accounts for the time sharing, and the factor of $2$ in the numerator of the SNR term is due to the fact that the user can transmit with twice the power since it only transmits half the time. 
Similarly, the ``ideal'' full-duplex ergodic capacity (the capacity when self-interference channel is zero) is easily shown to be
\begin{equation}
\label{eq:IdealFDDCap}	
C_{\rm Uplink}^{\rm Ideal FD} = \Ex \left[\log_2\left(1 + \frac 	{| h_S |^2 P_T}{N_R} \right)\right].
\end{equation}
Note that neither half-duplex nor ideal full-duplex capacities depend on $P_D$ or $P_R$. 


\ifLongAnalyticalDiscussion
Equations~(\ref{eq:FDCap})~(\ref{eq:HDCap})~and~(\ref{eq:IdealFDDCap}) are numerically evaluated in Figure~\ref{fig:CapPlot} to illustrate how passive suppression affects capacity. 
The values assumed for the system parameters, shown in Figure~\ref{tab:simParams}, are typical of WiFi systems.
The full-duplex capacity, $C_{\rm Uplink}^{\rm FD}$ is plotted as a function of the direct path strength, $P_D$, for several values of reflected path strength, $P_D$. 
Figure~\ref{fig:CapPlot} shows that as passive suppression reduces the direct path strength, $P_D$, the capacity grows until the direct path is attenuated to the level of the reflected paths.
However, Figure~\ref{fig:CapPlot} also shows that when reflection level is sufficiently low, passive suppression can enable significant full-duplex gains. 
For example, consider the anechoic chamber measurements for Antenna Configuration~\dirBigA\ shown in Figure~\ref{fig:suppressionChamber}. Without abortive shielding and cross-polarization, the direct path strength is $-45$~dB, but with abortive shielding and cross-polarization in place the direct path strength is suppressed to $-74$~dB. 
Plugging these values into Figure~\ref{fig:CapPlot} we see that employing absorptive shielding and cross-polarization would be the difference between achieving $4.7$ b/s/Hz and $7.9$ b/s/Hz: i.e., the difference between underperforming the half-duplex rate ($5.6$ b/s/Hz) by $15$\% and outperforming half-duplex by $30$\%. 

\else
Equations~(\ref{eq:FDCap})~(\ref{eq:HDCap})~and~(\ref{eq:IdealFDDCap}) are numerically evaluated in Figure~\ref{fig:CapPlot}. 
The values assumed for the system parameters, shown in Figure~\ref{tab:simParams}, are typical of WiFi systems. We see in Figure~\ref{fig:CapPlot} that once the direct path strength, $P_D$, is suppressed to the reflection level, $P_R$, the capacity no longer grows, which emphasizes the design implication of Result~1: that full-duplex infrastructure nodes should be deployed so as to neutralize nearby reflections.
\fi
}

\revision{
\subsection{Influence of Increased Frequency Selectivity on Active Cancellation}
In the above analysis we have assumed no limitation on the number of taps in the self-interference canceler. We now relax this assumption to study how increased frequency selectivity influences the performance of an active canceler with $N_{\rm Tap} \leq T_R$ taps. 
Denote the first $N_{\rm Tap}$ taps of $h_I$ as $h_I'[n] = h_I[0,1,\dots,N_{\rm Tap}-1]$.
We assume that the infrastructure node has perfect knowledge of $h_I'[n]$ and subtracts the known transmit signal, filtered by $h_I'[n]$, from its received signal forming
\begin{align}
y'[n] &= y[n] - h_I'[n]*x_I[n] \nonumber  \\
&= h_Sx_S[n] + \sum_{m=N_{\rm Tap}}^{T_R-1}h[m]x_I[n-m] \nonumber \\ 
&\ \ \  + h_I[n]*z_I[n] +  z_R[n] \label{eq:NTapCancelled}.
\end{align}
We define the average passive suppression, ${\alpha}_{\rm P}$, to be the average ratio of the transmit power to the power of the self-interference incident on the receiver. Similarly, we define the average active cancellation, ${\alpha}_{\rm A}$, to be the average ratio of the power of the self-interference incident on the receiver to the power of the self-interference after cancellation. 
It is easy to show from Equation (\ref{eq:NTapCancelled}) that the average passive suppression is  $\alpha_{\rm P} = \Ex \left[ \frac{1}{| h_I |^2} \right]$, and the average active cancellation is
\begin{equation}
\alpha_{\rm A} = \Ex \left[ \frac{| h_S |^2(P_T + N_T)}{\sum_{m=N_{\rm Tap}}^{T_R-1}|h_I[m]|^2 P_T  + \sum_{m=0}^{T_R-1}|h_I[m]|^2 N_T} \right],\label{eq:actCx} 
\end{equation}
where the expectations are with respect to the channel gain distributions specified above.
}

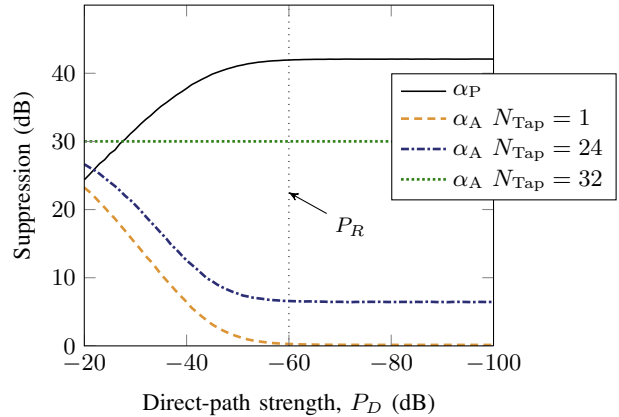
\begin{figure}[htdp]
\centering
\ifCLASSOPTIONonecolumn
{
\small
\begin{tikzpicture}
\begin{axis}[%
scale only axis,
width=0.30\textwidth,
height=0.25\textwidth,
xmin=-100, xmax = -20,
ymin=0, 
ymax=50,
xlabel={Direct-path strength, $P_D$ (dB)},
ylabel={Suppression (dB)},
x dir = reverse,
ylabel near ticks,
ytick = {0,10,...,40},
ylabel style={},
axis on top,
legend entries={
$\alpha_{\rm P}$,
$\alpha_{\rm A}\ N_{\rm Tap} =1$,
$\alpha_{\rm A}\ N_{\rm Tap} =24$, 
$\alpha_{\rm A}\ N_{\rm Tap} = 32$
},
legend style={at={(1.02,1.00)},anchor=north west,nodes=right}]
\node (reflection) at (axis description cs:0.65,0.35) {\footnotesize$P_R$};
\draw[->] (reflection) -- (axis description cs:0.5,0.45);
\draw[dotted] (axis description cs:0.5,0) -- (axis description cs:0.5,1);
\addplot [color=black,  line width=0.6pt]
table [x = directPathAtten_dB, y = Passive]{./DocData/Suppression.dat};
\addplot [color=KeynoteYellow,  line width=1.00pt, densely dashed]
table [x = directPathAtten_dB, y = Active1Tap]{./DocData/Suppression.dat};
%
\addplot [color=ECEpurple,  line width=1.00pt, densely dashdotted]
table [x = directPathAtten_dB, y = Active24Tap]{./DocData/Suppression.dat};
%
\addplot [color=KeynoteGreen,  line width=1.00pt, densely dotted]
table [x = directPathAtten_dB, y = Active32Tap]{./DocData/Suppression.dat};
%
\end{axis}
\end{tikzpicture}
}
\else
{
\small
\begin{tikzpicture}
\begin{axis}[%
scale only axis,
width=0.30\textwidth,
height=0.25\textwidth,
xmin=-100, xmax = -20,
ymin=0, 
ymax=50,
xlabel={Direct-path strength, $P_D$ (dB)},
ylabel={Suppression (dB)},
x dir = reverse,
ylabel near ticks,
ytick = {0,10,...,40},
ylabel style={},
axis on top,
legend entries={
$\alpha_{\rm P}$,
$\alpha_{\rm A}\ N_{\rm Tap} =1$,
$\alpha_{\rm A}\ N_{\rm Tap} =24$, 
$\alpha_{\rm A}\ N_{\rm Tap} = 32$
},
legend style={at={(0.75,0.8)},anchor=north west,nodes=right}]
\node (reflection) at (axis description cs:0.65,0.35) {\footnotesize$P_R$};
\draw[->] (reflection) -- (axis description cs:0.5,0.45);
\draw[dotted] (axis description cs:0.5,0) -- (axis description cs:0.5,1);
\addplot [color=black,  line width=0.6pt]
table [x = directPathAtten_dB, y = Passive]{./DocData/Suppression.dat};
\addplot [color=KeynoteYellow,  line width=1.00pt, densely dashed]
table [x = directPathAtten_dB, y = Active1Tap]{./DocData/Suppression.dat};
%
\addplot [color=ECEpurple,  line width=1.00pt, densely dashdotted]
table [x = directPathAtten_dB, y = Active24Tap]{./DocData/Suppression.dat};
%
\addplot [color=KeynoteGreen,  line width=1.00pt, densely dotted]
table [x = directPathAtten_dB, y = Active32Tap]{./DocData/Suppression.dat};
%
\end{axis}
\end{tikzpicture}
}
\fi
\caption{Passive suppression, $\alpha_{\rm P}$, and active cancellation, $\alpha_{\rm A}$, vs. direct-path strength with reflection strength of $P_R = -60$~dB and reflection extent of $T_R = 32$.  
	\label{fig:FreqSup}}
\end{figure}

\revision{
Equation~(\ref{eq:actCx}) is evaluated numerically in Figure~\ref{fig:FreqSup} for $N_{\rm Tap} = 1,24$ and $32$, where a reflection strength of $P_R = -60$~dB, a reflection extent of $T_R=32$, and parameter values shown in Figure~\ref{tab:simParams} are assumed. 
Figure~\ref{fig:FreqSup} emphasizes the design implication of Result~2. We see that when the passive suppression is low, a single-tap canceler can have good performance, but as the passive suppression increases and the channel becomes frequency selective, the single tap canceler degrades. Only if the self-interference canceler employs the full-number of taps in the channel, $T_R=32$, can the it continue to cancel significant amounts of self-interference in the presence of strong passive suppression.   
}



%
%
%

\section{Full-Duplex Prototype with Combined Passive and Active Suppression\label{sec:outdoor}}


 
We now investigate the performance of combined passive and active suppression in a prototype wideband OFDM full-duplex physical layer.

\subsection{Prototype Full-Duplex Infrastructure Node Description}

A full-duplex infrastructure node was prototyped using the WARPLab framework~\cite{WARPLab}, in which baseband signals are processed offline in MATLAB, then up-converted and transmitted over-the-air at 2.4 GHz in real time. The prototype implements a 20 MHz, 64-subcarrier OFDM PHY with a packet structure based on IEEE~802.11. The prototype infrastructure node was mounted outdoors on a 7~ft. pole just outside the Rice University Tudor Fieldhouse.
The prototype was evaluated using antenna Configuration \dirBigB~(Figure~\ref{fig:config_60_50}) with absorptive shielding, and with and without cross-polarization. Thus the mechanism combinations evaluated were $\sf{mech = \{DA, DCA\}}.$

For active self-interference cancellation, we adopted the per-subcarrier analog and digital cancellation mechanisms demonstrated in \cite{Duarte11FullDuplex, Sahai11RealTimeFD, Duarte2012FullDuplexWiFi}. 
Analog cancellation is implemented by estimating the self-interference channel via OFDM pilots, crafting a wideband cancellation waveform, and injecting the cancellation waveform into the receive signal path using an RF combiner, so that the the self-interference is cancelled prior to the RF front end.
%
Digital cancellation is implemented by estimating and subtracting, at baseband, the residual self-interference left after analog cancellation. Result~\ref{result:coherence} tells us that the coherence bandwidth of the self-interference channel will likely by high due to the passive suppression. Because the cancellation scheme is performed on a per-subcarrier basis, we expect that cancellation performance will be robust to such frequency-selectivity. 





\subsection{Protoype Evaluation Results}

\subsubsection{Total suppression exceeding 90~dB\label{sec:outdoor_suppression}}
\begin{figure*}[t]
\renewcommand{\subcapsize}{\small}
\centering
	\subfigure[CDF of self-interference suppression.\label{fig:CancellationCDF}]{
	\centering
	\begin{tikzpicture}
\pgfplotsset{ height=1.9in, width=2.2in}
\begin{axis}[%
scale only axis,
xlabel={Self-interference Suppression (dB)},
xlabel near ticks,
xmin = 55, xmax = 100, ymin = 0, ymax=1,
ylabel={Empirical CDF},
ylabel near ticks,
grid = major,
axis on top,
axis on top,
title = {\ },
legend entries={ 
	\small w/o Cross-pol: Passive, 
	\small w/ Cross-pol: Passive,
	\small w/o Cross-pol: Active,
	\small w/ Cross-pol: Active
},
legend style={ at={(0.5,-0.4)},anchor=center,nodes=right},
ytick = {0,0.25,0.5, 0.75, 1},
]
\addplot+[const plot, color=KeynoteBlue!50, mark = \empty, line width=1.5pt, mark size = 1.8pt, densely dashed]
	table [x = x, y = y]{./DocData/passiveCopolCDF.dat};
\addplot+[const plot,color=KeynoteRed, mark = \empty, line width=1.5pt, mark size = 1.8pt, densely dashed]
	table [x = x, y = y]{./DocData/passiveCrosspolCDF.dat};
\addplot+[const plot,color=KeynoteBlue!50, mark = \empty, line width=1.5pt, mark size = 1.8pt]
	table [x = x, y = y]{./DocData/activeCopolCDF.dat};
\addplot+[const plot,color=KeynoteRed, mark = \empty, line width=1.5pt, mark size = 1.8pt]
	table [x = x, y = y]{./DocData/activeCrosspolCDF.dat};
\end{axis}

\end{tikzpicture}
	 \vspace{60pt}
	}
	\subfigure[Percent uplink rate improvement of FD over HD.\label{fig:RateGain}]{
	\centering
	\begin{tikzpicture}
\pgfplotsset{ height=1.9in, width=2.2in}
\begin{axis}[%
scale only axis,
xlabel={Path Loss (dB)},
xlabel near ticks,
xlabel shift={-.25 em},
xmin = 84, xmax = 108,
ylabel={\% Improvement},
ylabel near ticks,
ylabel shift={-.75 em},
grid = major,
title={\ },
axis on top,
legend entries={\small Without cross-polarization, \small With cross-polarization},
legend style={ at={(0.02,0.05)},anchor=south west,nodes=right},
ytick = {-100,-50,...,100}
]
\vspace{30pt}
\addplot [color=KeynoteBlue!40, mark = diamond*, line width=1.00pt, mark size = 2.0pt]
	plot [error bars/.cd, y dir=both, y explicit,  
		error bar style={line width=1.00pt},
	  	error mark options={rotate=90,line width=1.00pt}
	]
	table [x = pathLoss, y = copolGain, y error = copolSTD]{./DocData/rateGain.dat};
\addplot [color=KeynoteRed, mark = *, line width=1.00pt, mark size = 1.8pt]
	plot [error bars/.cd, y dir=both, y explicit,  
		error bar style={line width=1.00pt},
	  	error mark options={rotate=90,line width=1.00pt}
	]
	table [x = pathLoss, y = crosspolGain, y error = crosspolSTD, ]{./DocData/rateGain.dat};

\end{axis}
%
\begin{axis}[%
yshift=-1.25cm,
xmin = 84, xmax = 108,
xlabel near ticks,
xlabel shift={-.25 em},
ymin = 1, ymax = 4,
hide y axis,
axis x line*=left,
scale only axis,
xlabel={Range (m) Shadowed Urban},
xtick = {84.8, 89.9, 95, 100.1, 105.2},
xticklabels = {70,100,150,215,315},
]
\end{axis}

\end{tikzpicture}
    }
\caption{Full-duplex prototype evaluation results.}
\end{figure*}
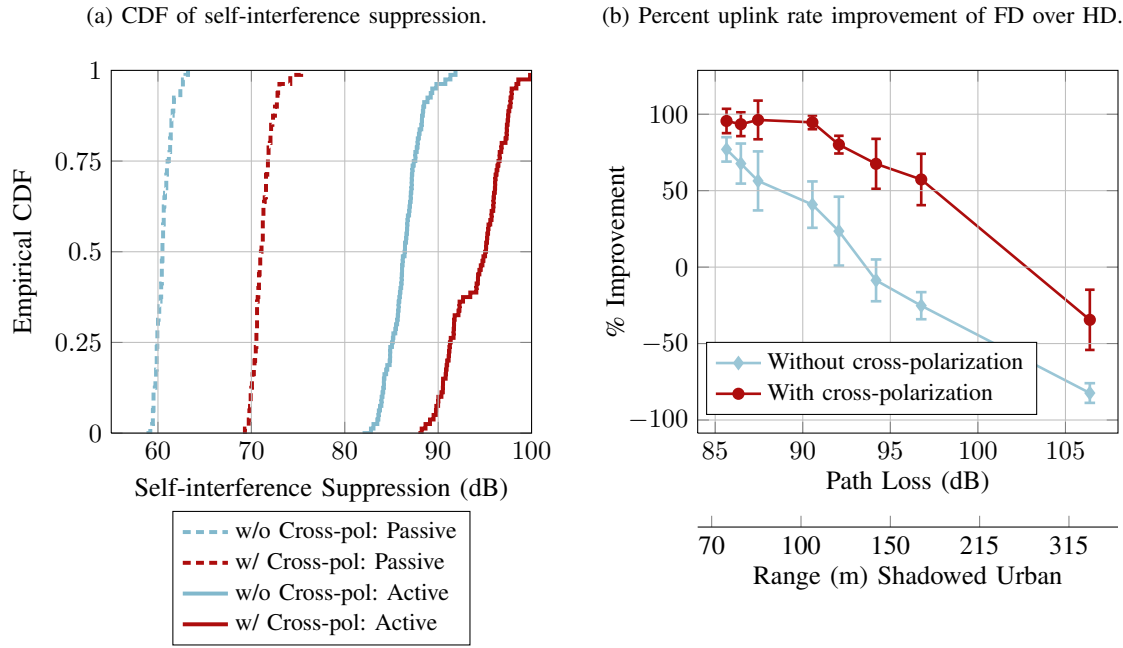

One-hundred packets were transmitted by the infrastructure node at a power of 7~dBm, and the amount of self-interference suppression was measured by observing the received signal strength indication (RSSI) from the WARP radio.
Figure~\ref{fig:CancellationCDF} shows the empirical CDFs for the self-interference suppression.  
We see in Figure~\ref{fig:CancellationCDF} that with directional isolation and absorptive shielding (${\sf mech = DA}$), 60~dB of passive suppression is achieved on the average. Adding cross-polarization (${\sf mech = DCA}$) increases the average passive suppression to 71~dB. These values are in line with the characterization of Section~\ref{sec:characterization}: since the outdoor deployment has weak reflections, we expect the achieved suppression to be near that seen in the anechoic chamber, in which we saw 72~dB of passive suppression for ${\sf mech = DCA}$.
Furthermore, we see in Figure~\ref{fig:CancellationCDF} that active cancellation adds around 25~dB of suppression, on the average, in both co-polarized and cross-polarized cases. 
In the cross-polarized case, the total self-interference suppression ranges from 87 dB to 100 dB, with 95~dB average suppression.\footnote{\revision{In the empirical CDF for ``w/ Cross-pol: passive $+$ active'', there is a slight flat spot at 95~dB. We believe this is an artifact of the finite-sampled (i.e. histogram) approximation of the distribution, and not evidence of a multi-moded distribution of self-interference suppression.}}
Thus the analog and digital cancellation mechanisms of~\cite{Duarte2012FullDuplexWiFi} are still able to provide 20-25 dB of self-interference cancellation, even after the self-interference is attenuated 70 dB by the passive mechanisms. 

%
%

\subsubsection{Full-duplex uplink-rate improvements at 100+ m} 
\begin{figure}[htbp]
\subfiguretopcapfalse
\centering
\subfigure[\Recip\ full-duplex: infrastructure node simultaneously transmits and receives to a single full-duplex user.]{
	\centering
	\ifCLASSOPTIONonecolumn
		\hspace{17pt}
		\includegraphics[width=0.42\textwidth]{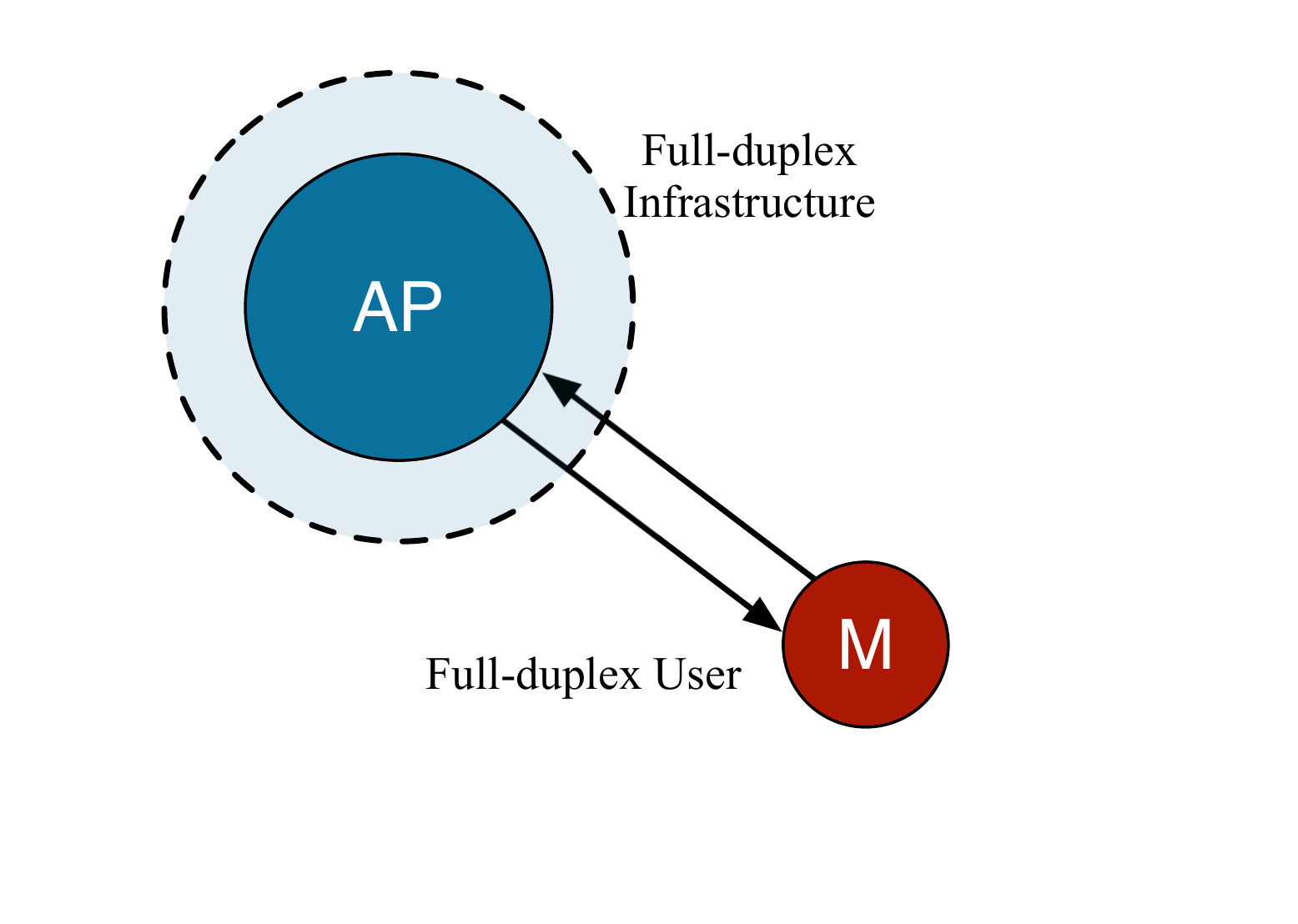}
	\else
		\hspace{22pt}
		\includegraphics[width=0.35\textwidth, clip=true, trim=0 2cm 0 0]{./DocGraphics/TwoNodeFD-eps-converted-to.pdf}
		\hspace{24pt}
	\fi
	\label{fig:recip}
}
\hspace{5pt}
\subfigure[\NonRecip\ full-duplex: infrastructure node receives from one user while transmitting to another user.]{
	\centering
	\ifCLASSOPTIONonecolumn
		\includegraphics[natwidth=449,natheight=328,width=0.42\textwidth]{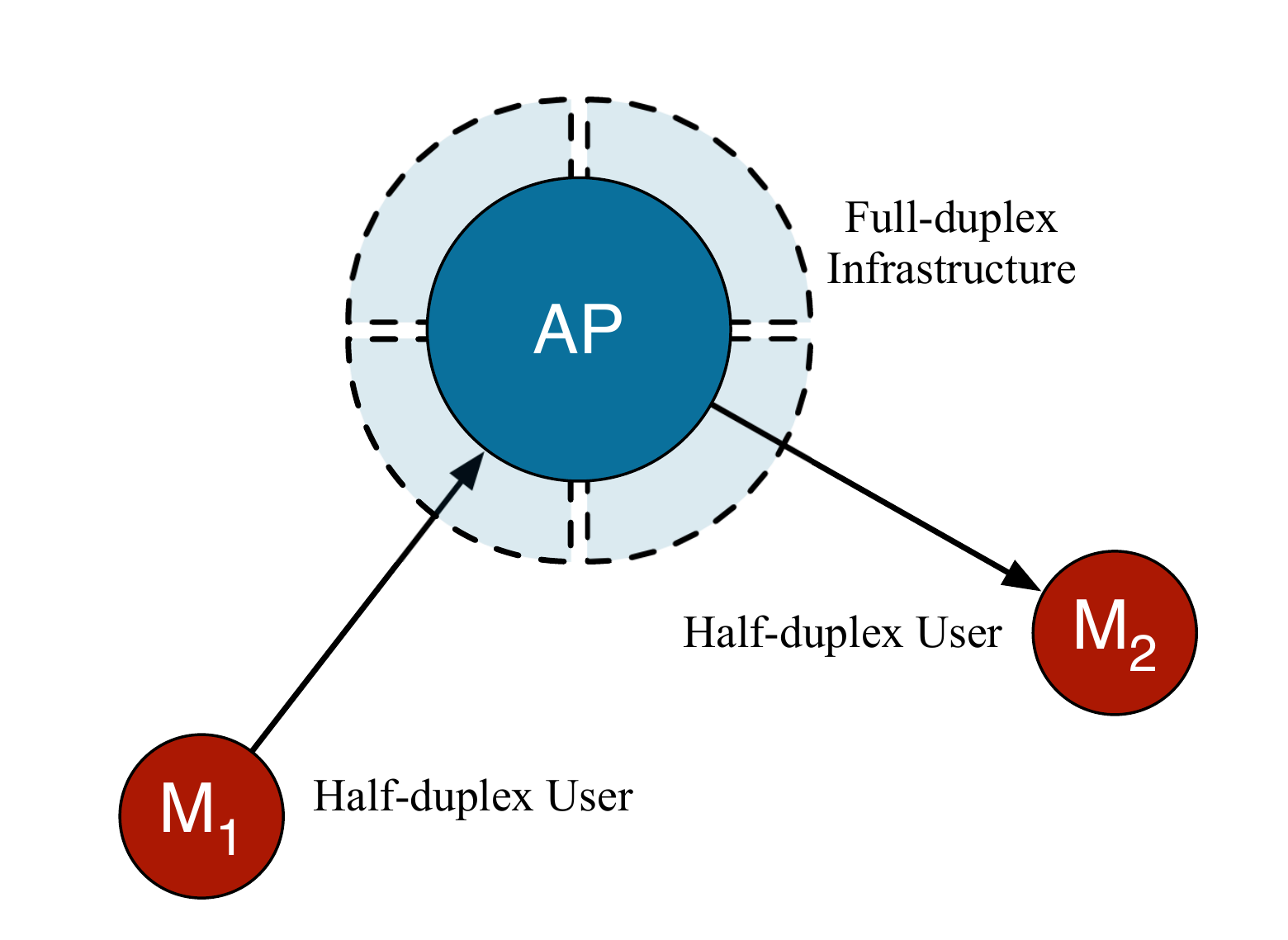}
	\hspace{17pt}
	\else
		\hspace{20pt}
		\includegraphics[natwidth=449,natheight=328,width=0.35\textwidth]{./DocGraphics/ThreeNodeFD-eps-converted-to.pdf}
		\hspace{20pt}
	\fi
	\label{fig:non-recip}
}
\caption{Scenarios for full-duplex transmissions.
}
\label{fig:FDmodes}
\end{figure}

\revision{
In a full-duplex infrastructure node with simultaneous uplink and downlink (such as a full-duplex relay or a full-duplex sectorized access~point as shown in Figure~\ref{fig:non-recip}), the uplink to the infrastructure node is the link that will suffer from self-interference. Thus a good system-level metric for comparing full-duplex performance against half-duplex is comparing the average achievable full-duplex uplink rate $\RFDUP$ to the comparable average achievable half-duplex uplink $\RHDUP$.  
To experimentally measure the average achievable rate, we employ the common practice of measuring the effective SNR (or SINR) for each packet, and computing the achievable rate for each packet from Shannon's formula \cite{CoverThomas} $R = \log(1+{\sf SNR})$. Averaging over all packets received, we obtain the average achievable rate. 
The effective SNR (SINR) of packet $p$ is measured by computing the average error vector magnitude, ${\sf \overline{EVM}}(p)$, which is the average distance between the received symbols (after equalization) in packet $p$ and the original transmitted symbols. Average EVM is used to find the effective SNR using the common conversion ${\rm SNR}(p) = 1/{\sf \overline{EVM}}(p)^2$ \cite{Mahmoud09EVMConversion}.

We measure the signal-to-self-interference-plus-noise ratio (SSINR) for the full-duplex uplink by having the infrastructure node receive packets from the uplink mobile node while simultaneously transmitting. From the EVM statistics we measure the $\SSINRFD$ for each packet.
Similarly, to measure the SNR of the comparable half-duplex uplink, we compute $\SNRUP$ from the EVM statistics of the packets received at the uplink, but without the infrastructure node transmitting simultaneously.
The average achievable rates of the full-duplex/half-duplex links are thus computed from the measured SSINR/SNR values using the formulae,
\begin{eqnarray}
	R_{\rm Uplink}^{\rm FD} &=& \frac{1}{N}\sum_{p = 1}^{N}\log_2[1+\SSINRFD(p)] \label{eq:FDRate} \\
	R_{\rm Uplink}^{\rm HD} &=& \frac{1}{N}\sum_{p = 1}^{N} \frac{1}{2}\log_2[1+\SNRUP(p)], \label{eq:HDRate}
\end{eqnarray}
where $p$ is the packet index, and $N$ is the total number of packets. Note that the pre-log factor of $\frac{1}{2}$ in Equation~(\ref{eq:HDRate}) is due to the assumption of even timesharing with downlink in half-duplex mode. 
To make the comparison fair in average power, the uplink mobile is given twice the transmission power (10~dBm rater than 7~dbm) in half-duplex mode, since we assume it defers to downlink half of the time. \revision{We note that the above method of characterizing a prototype's performance by computing achievable rates from measured SNR/SSINR is well-established in prior work such as \cite{Naren2010Beamforming, Duarte10Beamforming, Duarte11FullDuplex}.} For a very thorough explanation see Section~III of \cite{Duarte10Beamforming} and references therein.}




The achievable rates of the full-duplex uplink and comparable half-duplex uplink were measured as the range from of the uplink mobile to the infrastructure node was varied from 50 to 200 meters in order to encounter a variety of path losses. Figure~\ref{fig:RateGain} plots the percent improvement of the full-duplex uplink rate, $\RFDUP$, versus the half-duplex uplink rate, $\RHDUP$, as a function of the encountered path loss.\footnote{As expected, the encountered path loss was not monotonic with range due to shadowing effects, hence rate vs. range curves are noisy and difficult to interpret. Instead, we measured the encountered path loss for each mobile node location so that the performance could be indexed by this meaningful, repeatable parameter.} 
We see that at 86 dB path loss, the improvement over half-duplex is 86\% when directional isolation and absorptive shielding are applied together with active cancellation. When cross-polarization is also employed, the gain over half-duplex at 86~dB path loss is 96\%, meaning that the effect of self-interference is nearly eliminated, since the rate is nearly doubled. In Figure~\ref{fig:RateGain}, a second x-axis is also shown, which maps path loss to effective range in a typical urban channel~\cite{Rappaport}. With the best design, including cross-polarization, we expect to see full-duplex outperform half-duplex for up to 200~m range. Even without cross-polarization, the prototype full-duplex design outperforms half-duplex at effective ranges exceeding 100~m.  
Of particular note is the significant impact of cross-polarization. Figure~\ref{fig:CancellationCDF} showed that cross-polarization provides an extra 10~dB of total suppression, and we see the system-level impact of this extra 10 dB in Figure~\ref{fig:RateGain}: the same gains over half-duplex can be attained at 10~dB more path loss with cross-polarization than without. 

\revision{
\section{Discussion: Impact on the User's Access To the Infrastructure Node}
\label{sec:users}

There are two main scenarios in which an infrastructure node can leverage full-duplex capability to enhance its service to users: \recip\ full-duplex and \nonrecip\ full-duplex, illustrated in Figure~\ref{fig:FDmodes}.
Note that \recip\ full-duplex requires the user device to be full-duplex capable, while \nonrecip\ does not.
Due to the extra hardware burden that full-duplex would place on mobile devices, we expect the first applications of full-duplex will be \nonrecip\ scenarios, where the users are half-duplex, hence the emphasis on infrastructure nodes rather than user nodes in this paper. The interference from the uplink user to the downlink user in \nonrecip\ full-duplex can be managed by either scheduling non-interfering users as in \cite{Singh2011FDMAC}, or canceling the inter-user interference using advanced physical layers as in \cite{Bai12DecodeCancel}.

}

\revision{
\subsection{Impact of Directional Isolation}
\subsubsection{Supporting Full-duplex and Half-duplex Users}
We make the distinction between \recip\ and \nonrecip\ full-duplex, because directional isolation is often better suited for \nonrecip\ full-duplex applications.
For example, in Configuration~\dirBigA\ (Figure~\ref{fig:config_90_50}),
the two antennas cover disjoint areas, and cannot transmit and receive in the same direction, hence Configuration~\dirBigA\ is poorly suited to support \recip\ full-duplex transmissions. 
However, Configuration~\dirBigA\ is indeed well-suited to support \nonrecip\ full-duplex transmissions, when the uplink user is covered by one antenna the downlink user is covered by the other antenna as in Figure~\ref{fig:non-recip}. 
%
Configurations~\dirBigB~and~\dirSmall\ (Figures~\ref{fig:config_60_50}~and~\ref{fig:config_60_35}), both of which use $90\degree$ beamwidth antennas with $60\degree$ separation, also support \nonrecip\ full-duplex transmissions 
but can only support a \recip\ full-duplex transmission if the full-duplex-equipped user is in the $30\degree$ intersection of the two coverage patterns. Configurations~\omniBig~and~\omniSmall\ (Figures~\ref{fig:config_omni_50}~and\ref{fig:config_omni_35}) which use omnidirectional antennas support both \recip\ and \nonrecip\ full-duplex scenarios regardless of location of the user. 

\subsubsection{Managing directional access}
In a directional antenna system, control must be employed to ensure that any user is serviced by the antenna(s) with the strongest gain in the user's direction. 
Many algorithms have been proposed for adaptive directional antenna selection \cite{Ramanathan05DirectionalAdhoc,Ko2000DirectionalMAC,Choudhury02DirMAC}, and commercial access points with directional antennas that implement dynamic user selection are in the market \cite{XirrusSectoredBenefits,XirrusArrays}. It is outside the scope of this paper, but modest extension of the control algorithms used in these existing systems can be used to schedule concurrent \nonrecip\ full-duplex transmissions with directional antennas. 
}
\revision{
\subsection{Impact of Cross-polarization}

%
Cross-polarization at the infrastructure node can also affect users.  %
Assume the infrastructure node transmits with horizontal polarization and receives with vertical polarization.
The impact of cross-polarization on the user nodes depends on the environment as described below. 

\subsubsection{Non-line-of-environments}
In a non-line-of-sight environment, studies have shown that scattering leads to signals becoming \emph{de-polarized} \cite{Perez97PolModel}: the received wave is the superposition of many randomly polarized components. Thus the user is agnostic to the polarization state at the infrastructure node. A single-antenna user with any arbitrarily polarized antenna will be able to receive from the infrastructure node's horizontally polarized transmit antenna and transmit to the infrastructure nodes' vertically polarized receive antenna equally well.\footnote{The optimal design at the user device would have three orthogonally polarized antennas, combining power from all three so that all incident energy is captured. But this is not specific to full-duplex, but the case also for half-duplex in any non-line-of-sight environment.} 

\subsubsection{Line-of-sight environments}
In a line-of-sight environment, however, cross-polarization at the infrastructure node influences the polarization requirements of the user. If the user has only a single vertically polarized antenna, for instance, then it will have poor SNR when receiving the horizontally-polarized downlink. 
The ideal solution would be for users to match the infrastructure polarization: a half-duplex user with a dual-polarized antenna could transmit on the vertical port and receive on the horizontal port.
Dual-polarized antennas are available in form factors suitable for mobile phones \cite{Liu05DualPolPIFA}. 
The downside to polarization matching is that it requires the user device to either be in a stable orientation or to track its orientation. 
Smartphone sensors make tracking orientation feasible, and as orientation information was leveraged for antenna-pattern switching in \cite{Sani10DirectionalAntennas}, so it could be leveraged for polarization matching.
A simpler solution is for each user device to use a circularly polarized antenna, 
which couples with both vertical and horizontal polarization, but with a 3~dB loss due to polarization mismatch \cite{PozarWireless}. 
}

\section{Conclusion and Future Work}
\label{sec:discussion}
We have seen that a careful utilization of the passive self-interference suppression mechanisms of directional isolation, absorptive shielding, and cross-polarization can significantly improve the performance of full-duplex links, but some fundamental limitations are encountered. The amount of passive suppression is limited by reflected paths which are difficult to suppress passively. Moreover, because passive suppression eliminates the once-dominant direct self-interference path, the residual self-interference channel can be much more frequency selective. We have shown that a design that takes these capabilities and limitations into account can achieve a full-duplex link that outperforms a comparable half-duplex link even a ranges exceed 100~m. 

\revision{
A future step in this research direction is to develop a general theory for passive suppression. For example, given a size constraint, a coverage requirement, and a description of the scattering/reflection environment, one would like to solve for the antenna design and placement of absorbers which optimizes system capacity. Such a theory will require not only new analysis but new models, and we hope the results in this paper are a first step towards developing the models required for a general theory of passive suppression.  
}

\appendices

\section{Measurement Details and Uncertainty Analysis}
\label{sec:uncertainty}
\revision{
As mentioned in Section~\ref{sec:setup} the frequency response of the self-interference channel was measured using a network analyzer. We now discuss details how the measurement setup, and discuss the associated uncertainty. 

\subsection{Measurement Details}
A two-port network analyzer, such as the the Agilent E8364B \cite{PNA_E8364B} used in our study, measures the S-parameters of an arbitrary two-port RF network by sweeping a continuous wave at the input(s) and measuring the amplitude and phase of the output(s). 
In our case, the two ports of interest are the ports of the infrastructure node antennas, and $S_{21}$, the transmission coefficient, gives the amplitude and phase of the channel.
%
%
A full two-port calibration was performed with the Agilent N4691B \cite{CalN4691B} calibration kit connected \emph{after} the cables and connectors, thus correcting for any cable loss or coupling, and placing the phase and amplitude reference directly at the intended antenna ports. 
An intermediate frequency (IF) bandwidth of 20~kHz was used, as this provided a good balance between noise floor and frequency sweep time. To lessen the effect of noise, each frequency point was averaged over 20 repeated measurements.


\begin{figure}[htbp]
\begin{center}
\ifCLASSOPTIONonecolumn
		\includegraphics[natwidth=619,natheight=431,width=0.5\textwidth, clip=true, trim=0cm 0cm 0cm 0cm]{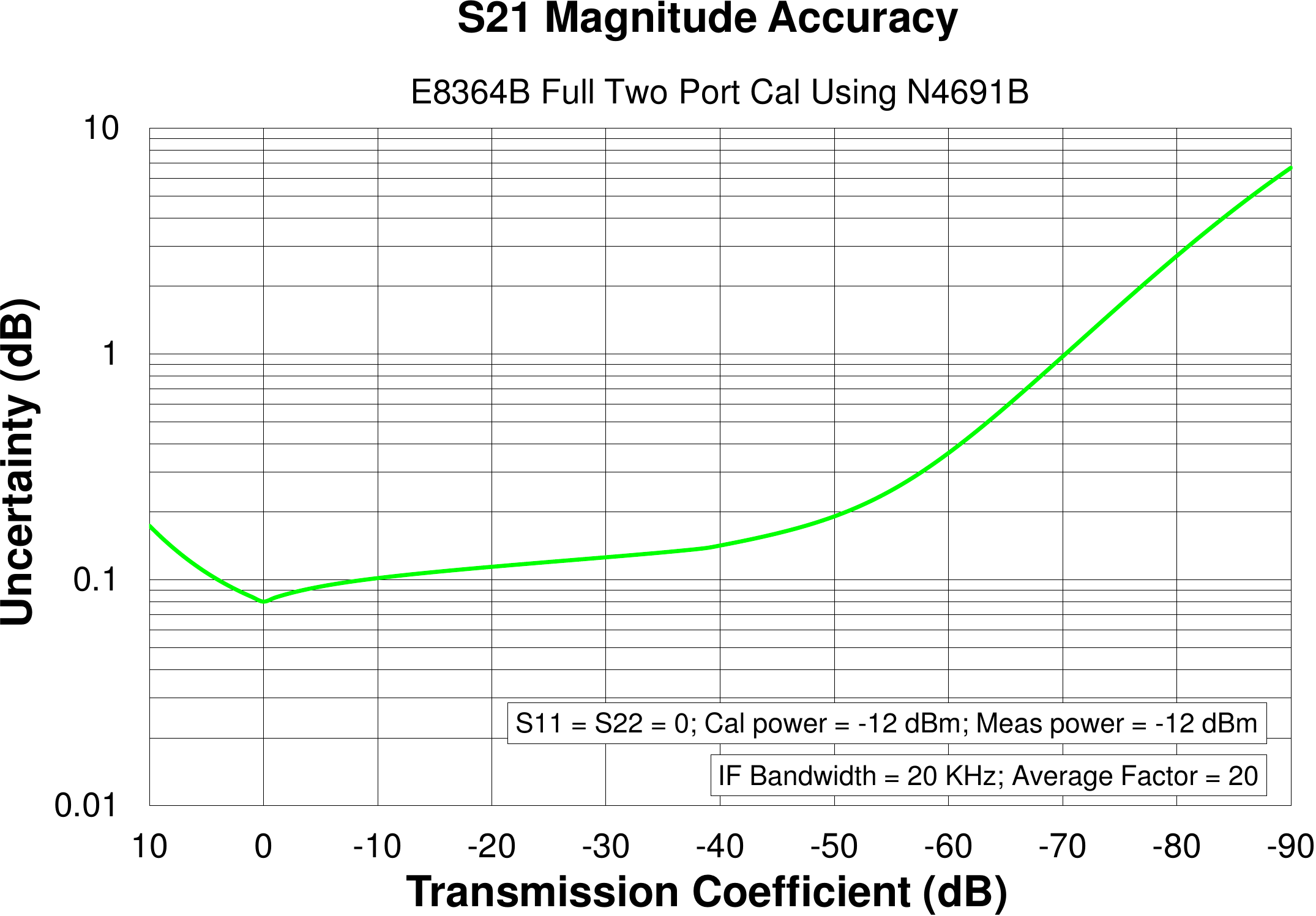}
	\else
		\includegraphics[natwidth=619,natheight=431,width=0.5\textwidth, clip=true, trim=0cm 0cm 0cm 0cm]{DocGraphics/UncertaintyPlot-eps-converted-to.pdf}
	\fi
\caption{Uncertainty curve for the network analyzer measurement setup presented in Section~\ref{sec:setup}}.
\label{fig:uncertainty}
\end{center}
\end{figure}

\subsection{Uncertainty Discussion}

Assessing uncertainty by merely computing empirical standard deviation over the 20 points measured would only quantify noise and not quantify the uncertainty due to imperfect calibration. 
Thankfully, Agilent provides a software tool called \emph{Uncertainty Calculator} \cite{UncertaintyCalulator}  
that computes the uncertainty associated with a measurement setup from the following key parameters: the IF bandwidth (20~kHz in our case) which affects the thermal noise, the averaging factor (20 in our case), and the calibration instrument used. \emph{Uncertainty Calculator} uses proprietary statistical characterizations of Agilent hardware to perform a root sum square uncertainty analysis and compute uncertainty as a function of the power of the quantity measured.  Figure~\ref{fig:uncertainty} plots the uncertainty curve generated for our setup, from which the uncertainty entries in Table~\ref{tab:Suppression} are taken.}

\ifCLASSOPTIONcaptionsoff
  \newpage
\fi

\bibliographystyle{IEEEtran}
\bibliography{IEEEabrv,/Users/evaneverett/Dropbox/Wireless_Literature/Bibliography/Research}

\newpage

%
\begin{IEEEbiography}[{\includegraphics[width=1in,height=1.25in,clip,keepaspectratio]{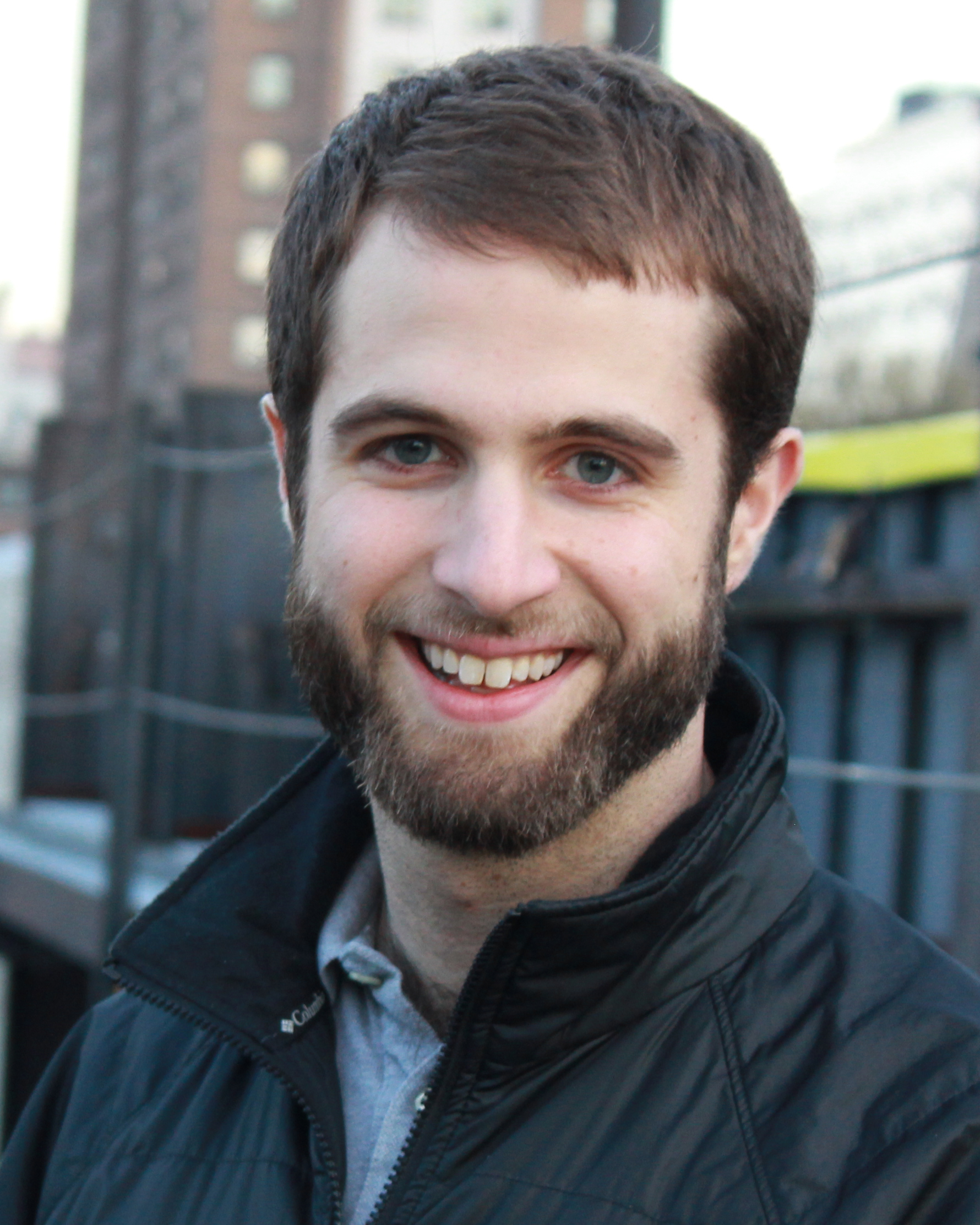}}]{Evan Everett} (S'06) received the B.Eng degree in Wireless Engineering and the B.S degree in Physics from Auburn University in 2010. He received the M.S. degree in Electrical Engineering from Rice University in 2012, where he is currently pursuing a Ph.D. His research focus is theory and design for full-duplex wireless communication. He received the 2010 Comer Award for Excellence in Physical Science from Auburn University, and he is a National Science Foundation Graduate Research Fellow.   \end{IEEEbiography}

\begin{IEEEbiography}[{\includegraphics[width=1in,height=1.25in,clip,keepaspectratio]{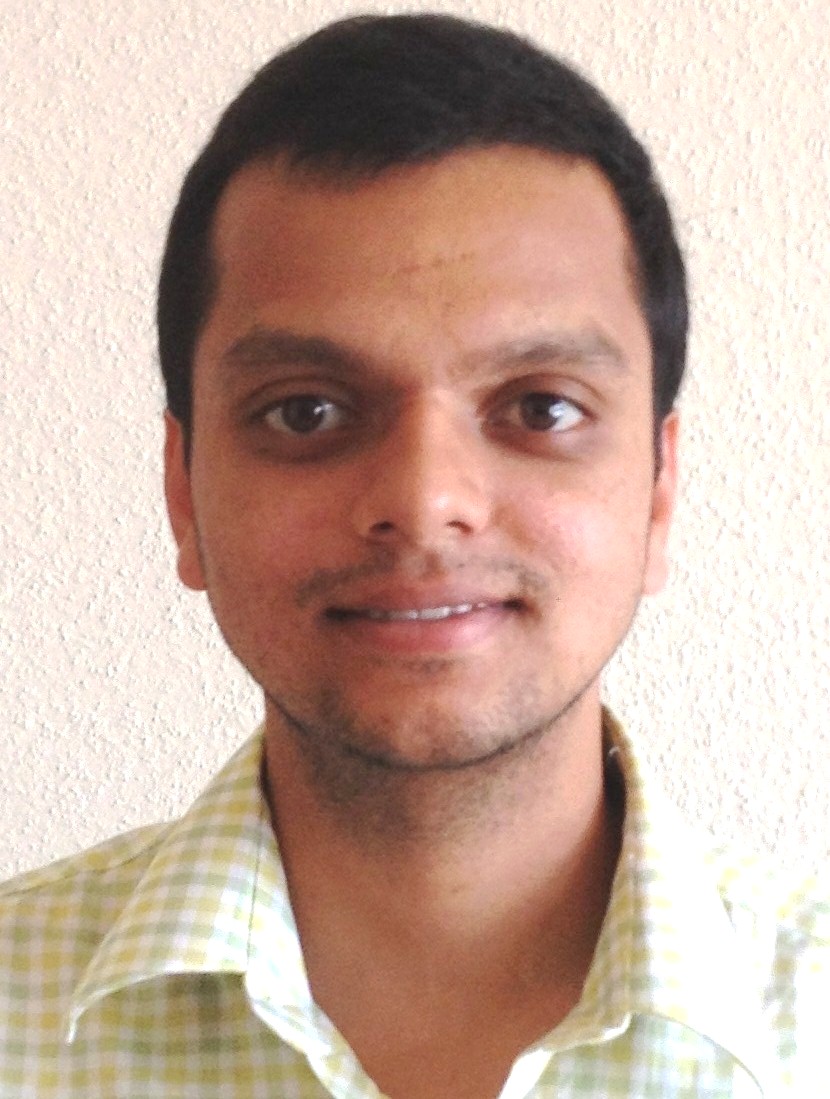}}]{Achaleshwar Sahai} received a Dual Degree (B.Tech
and M.Tech) in Electrical Engineering from the
Indian Institute of Technology Madras, Chennai in
2008. Currently he is pursuing a Ph.D. in Electrical
and Computer Engineering at Rice University, Houston,
TX. His research interests include information
theory and design of wireless communication systems.
\end{IEEEbiography}

\begin{IEEEbiography}[{\includegraphics[width=1in,height=1.25in,clip,keepaspectratio]{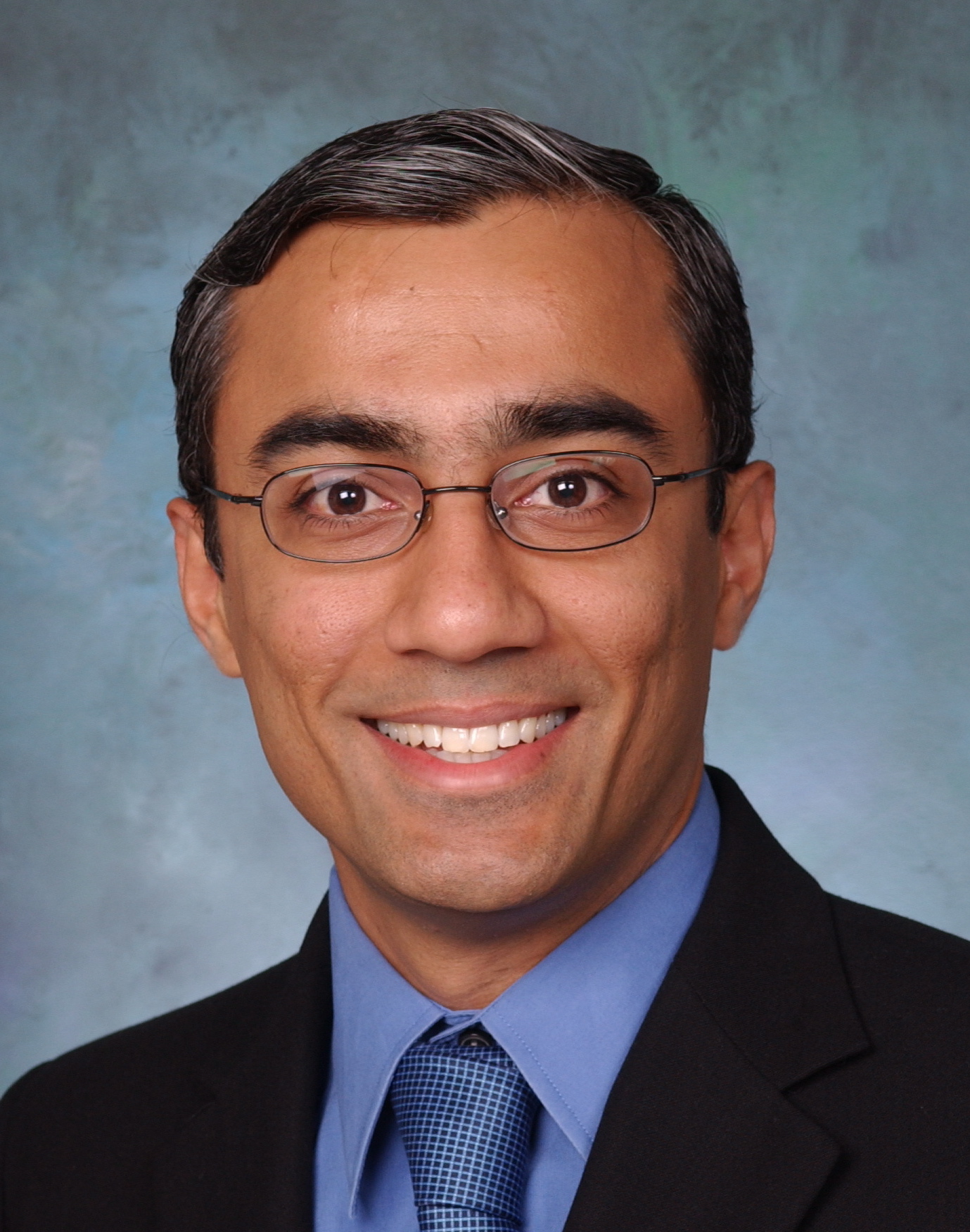}}]{Ashutosh Sabharwal} (S'91-M'99-SM'04) received the B. Tech. degree from the Indian Institute of Technology, New Delhi, India, in 1993 and the M.S. and Ph.D. degrees from The Ohio State University, Columbus, in 1995 and 1999, respectively. He is currently a Professor with the Department of Electrical and Computer Engineering, Rice University, Houston, TX. His research interests include information theory, communication algorithms and experiment-driven design of wireless networks. He received the 1998 Presidential Dissertation Fellowship Award. \end{IEEEbiography}

%
%
%




\end{document}